\documentclass[final,1p,times,twocolumn]{elsarticle}

\usepackage{amssymb}
\usepackage{amsmath}
\usepackage{booktabs}
\usepackage{graphicx} 
\usepackage{subcaption}
\usepackage{caption}   
\usepackage{siunitx}    
\usepackage{array,dsfont,multirow,threeparttable}      
\usepackage[ruled,vlined]{algorithm2e}

\begin{document}

\begin{frontmatter}



\title{Community-Level Modeling of Gyral Folding Patterns for Robust and Anatomically Informed Individualized Brain Mapping}
\author[a]{Minheng Chen} 
\author[a]{Tong Chen}
\author[a]{Yan Zhuang}
\author[a]{Chao Cao}
\author[a]{Jing Zhang}
\author[b]{Tianming Liu}
\author[c]{Lu Zhang$^*$}
\author[a]{Dajiang Zhu$^*$}
\affiliation[a]{organization={Department of Computer Science and Engineering, University of Texas at Arlington},
            city={Arlington},
            postcode={76019}, 
            state={TX},
            country={United States}}
\affiliation[b]{organization={School of Computing, University of Georgia},
            city={Athens },
            postcode={30602}, 
            state={GA},
            country={United States}}
\affiliation[c]{organization={Department of Computer Science, Indiana University Indianapolis},
            city={Indianapolis},
            postcode={46202}, 
            state={IN},
            country={United States}}
\begin{abstract}
Cortical folding shows substantial inter-individual variability yet contains stable anatomical landmarks that can support fine-scale characterization of cortical organization. Among these landmarks, the three-hinge gyrus (3HG) is a particularly informative folding primitive, exhibiting strong intra-species consistency alongside meaningful variations in morphology, connectivity, and functional relevance. However, prior landmark-based approaches typically model each 3HG in isolation, overlooking the fact that 3HGs form higher-order folding communities that capture mesoscale organizational structure. Ignoring this community-level organization oversimplifies gyral architecture and makes one-to-one landmark matching highly sensitive to fine-scale positional variability and noise.
We propose a spectral graph representation learning framework that explicitly models community-level folding units rather than isolated landmarks.
Each 3HG is characterized using a dual-profile representation integrating its topological surface context and structural connectivity fingerprint. A subject-specific spectral clustering module identifies coherent folding communities, followed by a topological refinement step that ensures anatomical continuity. To establish cross-subject correspondence, we introduce Joint Morphological–Geometric Matching (JMGM), which aligns community-level representations by jointly optimizing geometric and morphometric similarity.
Across more than 1,000 Human Connectome Project subjects, the resulting folding communities exhibit substantially reduced morphometric variance, stronger modular organization and superior cross-subject alignment, together with improved hemispheric consistency, compared to atlas-based and existing landmark- or embedding-based baselines.
These results demonstrate that community-level modeling of gyral landmarks provides a robust, anatomically grounded foundation for individualized cortical characterization, enabling more reliable correspondence and high-resolution subject-specific analyses.
\end{abstract}




\end{frontmatter}
\renewcommand{\thefootnote}{\fnsymbol{footnote}}
\footnotetext[1]{Corresponding author.
E-mail addresses: lz50@iu.edu (L. Zhang), dajiang.zhu@uta.edu (D. Zhu).}
\renewcommand{\thefootnote}{\arabic{footnote}}

\section{Introduction}
The human brain, a highly intricate and adaptive organ, serves as the epicenter of cognition, emotion, and behavior. Its outer layer, the cerebral cortex, is characterized by its highly convoluted structure, comprising convex gyri and concave sulci~\cite{rakic1988specification}.
This cortical folding pattern is fundamental to its function, as it allows a large surface area to be packed into a relatively small cranial space, thus maximizing the processing power of the brain~\cite{essen1997tension}. 
The human cerebral cortex is typically organized into multiple regions that subserve distinct functions, and understanding variability in the spatial layout and organization of these regions is essential for advancing our understanding of cortical organization, brain development, and disease-related alterations~\cite{lin2025cortical}.
To systematically study such region-specific organization and its inter-individual variability, brain mapping has emerged as a fundamental framework for identifying and characterizing the structural and functional regions of the brain~\cite{hagmann2008mapping}.
Precise brain mapping enables a better understanding of individual variability, inter-individual differences, and the impact of neurological conditions such as Alzheimer's disease~\cite{thompson2007tracking,apostolova2008mapping,xie2012mapping}, schizophrenia~\cite{callicott1998functional}, and epilepsy~\cite{hamandi2016non}. 
Over the years, many approaches have been proposed to map the brain, with parcellation techniques being one of the most widely adopted methods~\cite{fischl2002whole}. These techniques typically aim to segment the brain into distinct regions based on anatomical~\cite{destrieux2010automatic} or functional properties~\cite{craddock2012whole}, and these regions often form the basis for understanding brain connectivity and pathology~\cite{fan2016human}.
To further advance this direction, this work presents a novel cortical landmark-based system that establishes reliable community-level correspondences among gyral landmarks across individuals. Unlike conventional atlas-driven approaches, which force all brains to be mapped onto a common atlas and thereby overlook individual anatomical variability, our method preserves individual variability while enabling consistent cross-subject mapping. 
This innovation allows the construction of a finer-grained and anatomically grounded connectome that respects individual folding architecture.
Such a representation provides a more reliable substrate for individualized connectome analysis, allowing structural and connectivity patterns to be examined at a scale that is both biologically meaningful and consistent across subjects.

\begin{figure}[h!]

  \centering
  \centerline{\includegraphics[width=\linewidth]{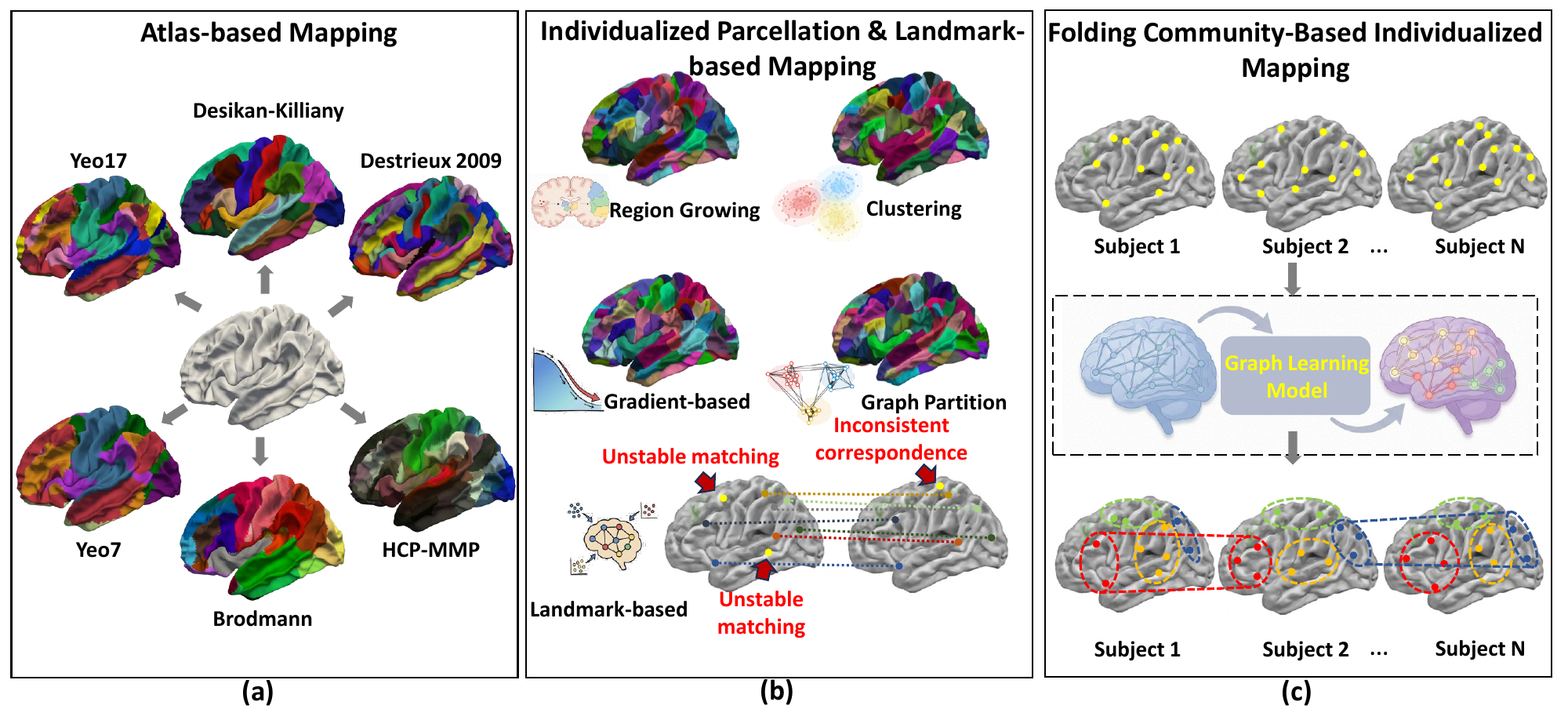}}
\vspace{-0.3cm}
\caption{Overview of current major paradigms for brain mapping and the proposed folding-community-based individualized framework.
(a) Atlas-based mapping assigns predefined regional labels via template-driven registration, enabling group-level comparison but often obscuring fine-scale folding variability.
(b) Data-driven individualized parcellation and prior landmark-based approaches aim to capture subject-specific patterns. However, data-driven parcellations often lack stable anatomical anchors for establishing cross-subject correspondence, while existing landmark-based methods overlook community-level folding organization and thus remain vulnerable to high inter-individual variability.
(c) Proposed folding community-based individualized mapping. Gyral landmarks are identified in each subject, organized into folding communities via graph learning, and aligned across subjects at the community level, yielding anatomically grounded and correspondence-ready folding representations.}
\label{fig:conceptual}
\end{figure}

Conventional brain mapping approaches predominantly rely on predefined templates or atlases to establish anatomical correspondence across individuals (Fig.~\ref{fig:conceptual}(a)). In this paradigm, individual brain scans are aligned to a common reference space through spatial registration and normalization, after which the cortex is divided into regions of interest (ROIs) according to atlas definitions~\cite{fonov2011unbiased,friston1995spatial}.
A representative example is the atlas-based pipeline implemented in FreeSurfer~\cite{destrieux2010automatic,desikan2006automated}. In this framework, a standard brain atlas is constructed from cortical surfaces that have been manually annotated according to established anatomical rules~\cite{fischl2004automatically}. For new subjects, FreeSurfer automatically assigns regional labels by aligning individual cortical surfaces to the atlas, extracting geometric information and local neighborhood relationships on the surface as features, and applying probabilistic modeling to determine regional assignments.
This procedure enables anatomically corresponding regions to be identified across subjects, facilitating efficient group-level analysis and the characterization of large-scale cortical organization at the population level.
However, these atlas-based approaches are fundamentally predicated on the assumption that individual brain anatomy, particularly the organization of gyri and sulci, closely resembles a common template shaped by shared genetic and environmental factors~\cite{sun2022genetic}.
This assumption becomes limiting in the presence of substantial inter-subject variability, particularly in cortical folding patterns, which exhibit pronounced differences in geometry and spatial organization across individuals.
As a result, rigid spatial alignment to a standard template may obscure subject-specific anatomical differences when establishing correspondence. Such limitations hinder the construction of reliable personalized brain maps and can reduce sensitivity to individual-specific anatomical organization, thereby limiting the generalizability of neuroimaging findings across diverse populations~\cite{akula2023shaping}. Consequently, there is a growing need for brain mapping strategies that better preserve individual anatomical variability while still enabling accurate and consistent cross-subject comparison.

To address the limitations of atlas-based brain mapping in capturing inter-individual anatomical variability, recent advances in computational methods have enabled a growing class of individualized brain mapping approaches (Fig.~\ref{fig:conceptual}(b)). By defining regions directly in each subject’s native space, these methods aim to preserve subject-specific organization while reducing reliance on a common template. Representative examples include clustering-based~\cite{han2020individualized,wang2015parcellating}, region growing-based~\cite{blumensath2013spatially}, and graph partition-based~\cite{honnorat2017sgrasp} individualized parcellation techniques, which delineate cortical subdivisions based on functional similarity or connectivity profiles.
Despite these advances, existing individualized parcellation methods exhibit several important limitations. Many approaches define region boundaries solely based on functional homogeneity~\cite{wang2015parcellating} or connectivity profiles~\cite{gordon2017precision}, often resulting in parcellations that lack direct anatomical interpretability. Consequently, the derived regions may not correspond to consistent cortical landmarks, limiting their ability to establish meaningful cross-subject correspondence.
Moreover, because functional homogeneity and connectivity profiles are themselves influenced by inter-individual differences in cortical morphology and folding organization, such parcellations are sensitive to anatomical variability across subjects~\cite{zhao2020functional}. This sensitivity makes it difficult to disentangle genuine subject-specific functional organization from variability indirectly induced by underlying anatomical differences.
Collectively, these limitations underscore the need for brain mapping frameworks that explicitly leverage stable anatomical landmarks to achieve both reliable individualized mapping and biologically meaningful comparability across subjects. Furthermore, despite their individualized nature, existing atlas-based and individualized approaches ultimately summarize cortical organization at the level of predefined regions, aggregating sub-ROI–level information into coarse ROI-to-ROI connectivity measures. Such aggregation discards fine-grained anatomical and connectivity details closely tied to local cortical folding, motivating the exploration of finer-scale, anatomically grounded representations~\cite{consagra2024continuous}.

Beyond parcellation-based strategies, accumulating evidence suggests that specific cortical folding features can serve as stable anatomical landmarks that anchor cortical organization across individuals~\cite{meng2014spatial}. 
Such landmarks reflect stable and consistent folding patterns embedded within the cortical architecture, shaped by neurodevelopmental constraints and exhibiting characteristic spatial relationships with surrounding cortical regions, making them well suited for individualized brain mapping.
Early work has demonstrated that folding-derived landmarks, such as sulcal pits and gyral peaks, exhibit high inter-individual reproducibility and close associations with functional specialization and developmental constraints~\cite{auzias2015deep,kaltenmark2020group,zhang2022gyral}. These findings established cortical landmarks as a promising alternative to region-based parcellation for capturing anatomically grounded individual variability.
Building on this insight, a line of research has sought to leverage cortical landmarks to construct individualized anatomical maps. 
However, in much of the existing literature, landmark identification relies on multimodal data, including fMRI, DTI, and multiple structural contrasts~\cite{zhu2013dicccol,lv2014holistic,jiang2016holistic,jiang2014anatomy,jiang2014intrinsic}. This reliance limits the applicability of such approaches to large cohorts and reduces the robustness, reproducibility, and scalability of landmark-based cross-individual mapping.
Within this broader class of anatomical landmarks, the three-hinge gyrus (3HG) offers a particularly compelling alternative, as it can be reliably identified from standard T1-weighted MRI while exhibiting a unique combination of strong intra-species stability and meaningful inter-individual variability~\cite{zhang2018exploring,li2017commonly,chen2017gyral}.
Prior studies have demonstrated that 3HGs possess distinctive morphometric signatures~\cite{li2010gyral}, denser and more heterogeneous structural connectivity~\cite{ge2018denser}, and hub-like roles in cortico-cortical communication~\cite{zhang2020cortical,he2022gyral}. 
Our recent work further shows that 3HG-based connectomes improve the diagnosis of neurodegenerative diseases and provide increased sensitivity to disease progression and staging~\cite{lyu2024mild,chen2025unified,chen2025representing}.
Despite these advantages, existing 3HG-based frameworks remain limited to single-landmark, node-level modeling. In these approaches~\cite{zhang2023cortex2vector,zhang2020identifying,huang2019multi,zhang2019group}, each 3HG is treated as an independent unit to be matched across individuals, with little consideration of the higher-order folding groupings that 3HGs naturally form. 
Recent work by Zhang et al.~\cite{zhang2023joint} has shown that such mesoscale folding communities carry meaningful anatomical structure, highlighting a critical gap in current correspondence methods. Ignoring this community-level organization oversimplifies gyral architecture and renders correspondence highly sensitive to fine-scale positional variability, inter-individual differences in folding, and noise inherent to MRI data.
More fundamentally, the brain is an intrinsically networked system, in which structure and function emerge from coordinated interactions among groups of elements rather than from isolated nodes~\cite{meunier2010modular,betzel2017multi}. From this perspective, community-level representations provide a more biologically meaningful and principled description of cortical organization than single-landmark matching. As a result, one-to-one landmark correspondence becomes inherently unstable and error-prone, motivating a shift from isolated landmark matching toward community-level modeling and alignment to support more robust and accurate cross-subject anatomical mapping.

In this paper,  we introduce a novel approach to individualized brain mapping by employing spectral graph representation learning of gyral folding patterns (Fig~\ref{fig:conceptual}(c)).
To characterize individual 3HG landmarks, we first developed a dual‑profile feature representation that integrates each landmark’s topological context with its structural connectivity fingerprint. Building on these enriched features, we designed a subject‑specific, deep‑learning–driven graph spectral clustering algorithm to partition each subject’s 3HGs into a predetermined number of communities. 
To ensure anatomical plausibility, a subsequent connection‑correction module enforces topological consistency among clustered 3HGs. Using this pipeline, we then established robust cross‑individual correspondences at the community level by using a Joint Morphological-Geometric Matching (JMGM) strategy.
We validated our framework on the Human Connectome Project dataset, and through extensive qualitative and quantitative analyses, we demonstrated that it not only yields stable within‑subject 3HG partitions but also preserves community‑level consistency across subjects. 
Collectively, this work establishes a rigorous, anatomically grounded community-level correspondence framework for 3HG cortical landmark, serving as an essential prerequisite for subsequent individualized brain‑mapping efforts.

\section{Methodology}
\begin{figure*}[h!]

  \centering
  \centerline{\includegraphics[width=\linewidth]{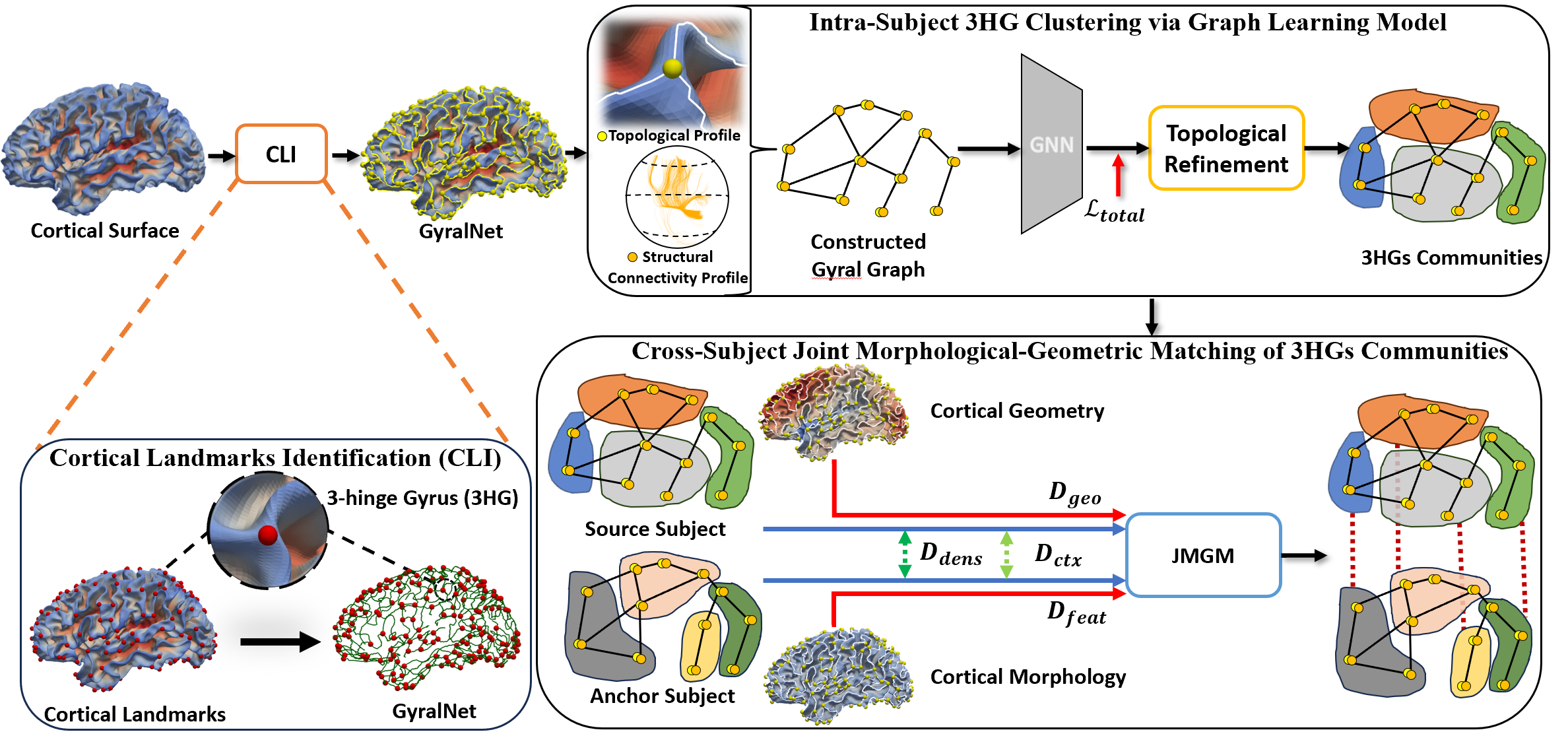}}
\vspace{-0.3cm}
\caption{Overview of the proposed pipeline for individualized cortical landmark community-level correspondence. 3HGs are identified on cortical surface, then clustered via subject-specific GNNs with topological refinement to ensure spatial continuity. Cross-subject correspondence of 3HGs communities is obtained through joint morphological–geometric matching, solved with a Hungarian assignment and anchored for consistent cohort-wise labeling.}
\label{fig:main}
\end{figure*}
\subsection{Overview of the Individualized Cortical Landmark Correspondence Construction Pipeline}
This study aims to construct an individualized cortical landmark correspondence by analyzing the local fine-scale gyral folding pattern, \textit{i.e.}, 3HGs. Our method is designed to capture the anatomical variability of gyral folding across subjects using spectral graph representation learning. As shown in Fig.~\ref{fig:main} , the pipeline consists of four main components:
1) a dual-profile feature representation for each 3HG based on its topological context and structural fiber connectivity pattern;
2) a subject-specific clustering step using graph neural networks to identify coherent 3HGs communities within a hemisphere;
3) a connectivity-constrained topological refinement process to ensure spatial continuity and anatomical plausibility of clusters; and
4) a cross-subject mapping procedure to establish correspondence between clusters across individuals by using a Joint Morphological-Geometric Matching strategy.
This framework enables the construction of consistent, anatomically grounded cortical maps that reflect subject-specific folding patterns and support population-level analysis. And the details of each component will be introduced in the following sections.
\subsection{Feature Representation of the 3-Hinge Gyrus}
Precisely characterizing the 3HGs within individual subjects through feature representations that capture their topological structure and connectivity patterns, while ensuring that each 3HG is uniquely defined in each subject and enabling consistent anatomical correspondence across individuals, presents a significant challenge in the cross-subject alignment of gyral folding patterns.
Our previous work, cortex2vector~\cite{zhang2023cortex2vector}, leverages brain region annotations provided by a specific cortical atlas to construct hierarchical representations of the 3HG. This was achieved by encoding the location of each 3HG point and its neighboring nodes based on the atlas-defined brain regions they occupy. A recent study~\cite{chen2025using} further improved this encoding method by introducing the structural similarity between nodes to better ensure the intra-subject distinctiveness of each 3HG.
Although existing methods succeed in assigning higher similarity to nodes that share topological properties, they remain susceptible to individual variability when used to establish correspondences among 3HGs across subjects. This susceptibility arises because a single subject may exhibit multiple 3HGs with comparable topology, or a given 3HG on one cortical surface may lack any counterpart with similar topological features in another subject.

To overcome these limitations, we introduce a dual-profile feature representation that captures both the topological and structural characteristics of each 3HG.
The topological profile encodes intrinsic topological connectivity and neighborhood relationships of the 3HGs, ensuring that nodes with similar folding configurations and spatial locations are grouped together. 
Specifically, we use the one-hop structural similarity of 3HGs described in~\cite{chen2025using} as the topological profile.
The structural connectivity profile is built upon the trace-map model~\cite{zhu2011discovering, ZHU2012optimization}, a quantitative fiber bundle descriptor that characterizes the morphology of the fibers traversing each 3HG.
Consequently, for each 3HG we derive two feature vectors, one describing its topological similarity profile and the other describing its structural connectivity profile.
We concatenate these two vectors together to form the node feature input for the GNN model which will be introduced later. 
\subsection{Intra-Subject 3HG Clustering via Graph Neural Networks}
Based on the extracted dual-profile features, we formulate the clustering of 3HGs within a single cerebral hemisphere as a node clustering task on a subject-specific graph. Each node represents a 3HG, and the edges encode topological adjacencies obtained by GyralNet~\cite{chen2017gyral}, a cortical gyri organization pattern. We employ a two-layer graph neural network (GNN) to learn node embeddings that integrate both local features and graph context, enabling the unsupervised partitioning of the graph into spatially coherent and anatomically meaningful clusters. This step produces an individualized parcellation of the cortical folding patterns, tailored to each subject’s unique folding topology.

Given $n$ 3HGs on a cortical hemisphere, the adjacency matrix $\textbf{A}\in \mathbb{R}^{n\times n}$ obtained from GyralNet, and the dual-profile features $\textbf{X}\in \mathbb{R}^{n\times r}$ corresponding to these 3HGs. Our objective is to compute a soft assignment matrix $\mathcal{P}\in \mathbb{R}^{n \times k}$ that divides these 3HGs into $k$ clusters.
We first apply a two-layer graph convolutional network to compute the node embedding $\textbf{Z}\in \mathbb{R}^{n \times d}$, where $d$ is the hidden dimension. For the 
$t$-th layer of the network, the inputs are the feature matrix $\textbf{X}^{(t)}$ produced by the previous layer and the adjacency matrix $\hat{\textbf{A}}$ ($\textbf{X}^{(1)}=\textbf{X}$). The layerwise propagation can be written as:
\begin{equation}
X^{(t+1)} = \sigma\bigl(\hat{D}^{-1/2}\,\hat{\mathbf{A}}\,\hat{D}^{-1/2}\,\mathbf{X}^{(t)}\,\mathbf{W}^{(t)}  + \mathbf{X}^{(t)} \mathbf{W}^{(t)}_{\text{skip}}\bigr)
\end{equation}
where 
$\hat{\mathbf{A}} = A + I$, $\hat{D}_{ii} = \sum_{j}\hat{\mathbf{A}}_{ij}$, $\hat{D} = \mathrm{diag}(\hat{D}_{ii})$, $\sigma$ is an element-wise activation function, $\textbf{W}^{(t)}$ and $ \mathbf{W}^{(t)}_{\text{skip}}$ are the learnable weight matrices at layer $t$.
And the output node embeddings pass through a Multi-Layer Perception layer and a softmax function to obtain the soft assignment probability $\mathcal{P}\in \mathbb{R}^{n\times k}$ (shown in Eq.~\ref{eq::network}), which makes the assignment process differentiable.
\begin{equation}
\label{eq::network}
  \mathcal{P}= \text{softmax}(\text{MLP}(\textbf{Z}))
\end{equation}

In our initial study~\cite{zhuang2025gyralnet}, we take maximizing the spectral modularity~\cite{tsitsulin2023graph} of the GyralNet as the cost function of this unsupervised graph clustering task. Graph modularity~\cite{newman2006modularity} quantifies the strength of a network division into communities by comparing the observed intra‐community edge density to that expected under a random null model, and it has been widely adopted as the objective function for graph clustering~\cite{tsitsulin2023graph,liu2024revisiting,bhowmick2024dgcluster}. The loss function for graph spectral modularity maximization can be expressed as follows:
\begin{equation}
\label{eq::modularity}
  \mathcal{L}_m=
- \frac{1}{2m} \operatorname{Tr}(\mathcal{P}^{\top} \mathbf{B} \mathcal{P})
+ \frac{\sqrt{k}}{n} \left\| \sum_{i} \mathcal{P}_i^{\top} \right\|_F - 1
\end{equation}
However, relying solely on the maximization of spectral modularity does not yield an optimal partitioning. First, this heuristic overlooks the feature attributes of individual 3HG nodes. Second, as our prior work~\cite{zhuang2025gyralnet} has shown, it lacks sensitivity to smaller communities and fails to ensure topological continuity within each community. To address these shortcomings, we introduce several auxiliary optimization objectives that serve as loss terms in our graph neural network clustering framework.

As in many previous works on constructing brain networks~\cite{van2008normalized,mezer2009cluster,craddock2012whole}, we need to maintain the spatial continuity of each cluster to distinguish between network nodes and large-scale networks of nodes. 
To achieve this,  we first compute the matrix $M \in \mathbb{R}^{n\times n}$ of shortest path distances between all node pairs in the graph.
For each cluster $c$, we construct a weight matrix $\mathcal{P}_{c}\mathcal{P}_{c}^{\top}$ through the soft assignment of the node $\mathcal{P}_{c}$  and calculate the weighted average of the distances of all nodes in the cluster.
We then introduce a loss term that penalizes large distances among nodes assigned to the same cluster, thereby encouraging the model to group together nodes that are separated by small graph distances.
The loss function $\mathcal{L}_{p}$ is shown below:
\begin{equation}
    \mathcal{L}_{p} = \frac{1}{k} \sum_{c=1}^{k} \frac{\sum_{i=1}^{n} \sum_{j=1}^{n} \mathcal{P}_{i,c} \mathcal{P}_{j,c} M_{ij}}{\sum_{i=1}^{n} \sum_{j=1}^{n} \mathcal{P}_{i,c} \mathcal{P}_{j,c}}
    \label{eq::shortest path}
\end{equation}
The graph Laplacian provides a concise representation of network connectivity, with the multiplicity of its zero eigenvalues corresponding to the number of connected components. Since directly counting zero eigenvalues is not differentiable, so we instead minimize the second smallest eigenvalue to encourage each cluster to form a single connected component. However, this objective proves unstable in practice and can lead to vanishing gradients during training. To overcome this issue, we introduce a relaxation that maximizes the determinant of the Laplacian matrix for each cluster, providing a smoother and more numerically stable surrogate for enforcing topological continuity:
\begin{equation}
    \mathcal{L}_{\text{det}} = -\log \det (\mathcal{P}^\top \Theta \mathcal{P} + J)
    \label{eq:det_equation}
\end{equation}
where $\Theta = \hat{D}-\hat{\textbf{A}}$ is the graph Laplacian of the GyralNet and $J = \frac{1}{k} \mathbf{1}_k \mathbf{1}_k^\top $.
Through these two terms, we ensure global topological continuity and intra-cluster structural homomorphism.

Ensuring that nodes within the same community exhibit homogeneous features while nodes in different communities display heterogeneous features is essential for meaningful subnetwork delineation.
Contrastive learning is frequently employed in graph clustering to enforce consistency of representations among nodes in a cluster~\cite{liu2024revisiting,devvrit2022s3gc,liu2024reliable}.
Guided by anatomical priors of cortical parcellation, we posit that 3HGs located in adjacent cortical regions should have similar embedding features. To formalize this, we define a binary matrix $C\in \{0,1\}^{n \times n}$ such that $C_{ij} = 1$ if node $i$ and node $j$ form a positive pair (indicating regional adjacency and feature similarity), and $C_{ij} = 0$ otherwise.
Node similarity matrix is obtained from the node embedding $\textbf{Z}\in \mathbb{R}^{n\times d}$, $S=\frac{\textbf{Z}\textbf{Z}^\top}{\tau}\in \mathbb{R}^{n \times n}$, where $\tau$ is the temperature hyperparameter.
We use an InfoNCE~\cite{oord2018representation,zhang2025spineclue} style loss to encourage nodes with predefined similarity to have similar embeddings in the latent space:
\begin{equation}
    \mathcal{L}_c = \frac{1}{\sum_i \mathds{1}_{| \Omega(i) | > 0}} \sum_{i=1}^{N} \mathds{1}_{| \Omega(i) | > 0} \cdot-\log \left( \frac{\sum_{j \in \Omega(i)} \exp(S_{ij})}{\sum_{k=1}^{n} \exp(S_{ik})} \right)
    \label{eq:contrastive}
\end{equation}
where $\Omega(i) =\{j | C_{ij}  =1\}$ is the set of positive nodes for anchor $i$.
The overall learning objective for 3HG clustering is:
\begin{equation}
    \mathcal{L}_{total}= \mathcal{L}_m+\alpha\mathcal{L}_{p}+\beta \mathcal{L}_{\text{det}}+\gamma \mathcal{L}_c
    \label{eq:loss}
\end{equation}
where $\alpha$, $\beta$ and $\gamma$ are hyperparameters trading off the loss functions respectively.
It is important to note that our GNN-based clustering approach operates within individual subjects. For each cortical hemisphere of a given subject, the algorithm iteratively groups the 3HGs on GyralNet until the loss function converges, and the resulting assignment is taken as the final clustering. Because cortical folding patterns vary uniquely across individuals, a separate model is trained from scratch for each subject’s hemisphere.
\subsection{Connectivity-Constrained Topological Refinement of Cortical Clusters}
The initial clustering results may contain artifacts such as spatial discontinuities, non-contiguous regions, or small fragmented patches. To address these issues, we apply a topological refinement procedure that enforces intra-cluster connectivity and smooths cluster boundaries. Specifically, disconnected components within each cluster are identified and either merged with adjacent dominant clusters or relabeled based on minimum cut criteria. This refinement step ensures that the resulting cortical parcels are anatomically plausible and suitable for downstream analysis. 
Algorithm \ref{algo::refinement} presents the pseudocode for the connectivity-constrained refinement step.

Our refinement strategy consists of the following steps: 1) First, for each cluster, if it is not fully connected, the algorithm will retain the largest connected component and mark the remaining components as to be redistributed.
If no extra connected components are detected in any cluster, and the number of non-isolated clusters is still below the preset value $k$, the algorithm will select the current largest cluster, build a weighted graph based on node embedding on the subgraph corresponding to the cluster, and the weight of the edge of the graph is equal to the cosine similarity of the node embeddings of the two nodes, \textit{i.e.} for node $i$ and node $j$, the edge weight between them is $1-cos(\textbf{Z}_i,\textbf{Z}_j)$.
We then apply the spectral clustering method, \textit{i.e.}, normalized cut spectral clustering algorithm ~\cite{shi2000normalized} to this cluster to split it into two connected subclusters due to its robustness to outliers and favorable performance when compared to other clustering methods~\cite{shen2010graph}.
2) After that, the strategy enters the iterative correction phase: in each iteration, for each non-isolated node with only one neighbor, if its only neighbor belongs to a different cluster from itself, the node is merged into the neighbor cluster to eliminate the break of a single node across clusters;
at the same time, for clusters that still have multiple connected components, the components are reassigned as a whole to the optimal candidate cluster by examining the similarity of the node embeddings between the external neighbors of the component nodes and the component nodes. After iterating until the cluster labels are stable, we obtain the final cluster assignment result after refinement.
\begin{algorithm}[H]
\caption{Connectivity-Constrained Refinement of Cortical Surface Clusters}
\KwIn{
    Soft cluster assignments $\mathcal{P}\in\mathbb{R}^{n\times K}$;
    surface graph adjacency $A$;
    node features $\mathbf{Z}$;
    target cluster count $K$.
}
\KwOut{
    Final hard labels $\Omega\in\{0,\dots,K-1\}^n$.
}

\textbf{Initialization.}\\
Construct graph $G$ from $A$;  
Initialize hard labels via $\Omega_i=\arg\max_j \mathcal{P}_{ij}$;  
Mark isolated vertices with $\Omega_i=-1$.

\medskip
\textbf{Enforcing Cluster Count.}\\
\While{number of valid clusters $<K$}{
    \ForEach{label $q$}{
        Identify connected components of $G[q]$; \\
        \If{multiple components exist}{
            Assign a new label to one component and continue;
        }
    }
    Select the largest cluster $q^\star$; \\
    Apply spectral bipartitioning on $G[q^\star]$ using features $\mathbf{Z}$; \\
    Assign a new label to the smaller partition.
}

\medskip
\textbf{Connectivity Refinement.}\\
\Repeat{until convergence}{
    \ForEach{vertex $i$}{
        \If{$i$ is a leaf node and $\Omega_i\neq\Omega_{\text{nbr}(i)}$}{
            Propagate the neighbor label to $i$;
        }
    }
    \ForEach{cluster $q$}{
        Extract connected components of $G[q]$; \\
        \ForEach{non-dominant component $Q$}{
            Select the best target cluster $q^\star$ based on \\ 
            \hspace{2.5em}(i) boundary edge density and  
            \hspace{0.5em}(ii) mean cosine similarity; \\
            Reassign all nodes of $Q$ to $q^\star$.
        }
    }
}

\medskip
\textbf{Final Step.}\\
Normalize cluster indices into $\{0,\dots,K-1\}$.
\label{algo::refinement}
\end{algorithm}

\subsection{Constructing Cross-Subject Correspondence of 3HG Clusters via Joint Morphological-Geometric Matching}

Let each hemisphere of a subject contain $K$ clusters of 3HGs detected by our pipeline. 
For a given subject, denote by $\{\Omega_i\}_{i=1}^K$ the cluster sets on the unit sphere $\mathbb{S}^2$ and by $\mathbf{u}_p\in\mathbb{S}^2$ and $\mathbf{f}_p\in\mathbb{R}^F$ the spherical coordinate and the multi-modal morphological feature vector (including curvature, sulcal depth, and surface area derived from the white surface) at vertex $p$, respectively.
To robustly represent local morphology without relying on a single vertex, each cluster $\Omega_i$ is expanded to a topological $k$-ring neighborhood $\mathcal{N}_i^{(k)}$ on the spherical mesh, from which we uniformly subsample up to $M$ vertices.
This yields two sets per cluster:
\begin{equation}  
\mathbf{U}_i=\{\mathbf{u}_p\}_{p\in \mathcal{N}_i^{(k)}}, 
\qquad
\mathbf{F}_i=\{\mathbf{f}_p\}_{p\in \mathcal{N}_i^{(k)}}.
\end{equation}
We compare clusters via distances between the \emph{empirical feature distributions} supported on $\mathbf{F}_i$ rather than simple summary statistics.
Specifically, for two subjects $a$ and $b$ and clusters $i$ and $j$, we define
\begin{equation}
D_{\mathrm{feat}}(i,j)
\;=\;
\mathsf{SW}_\varepsilon\big(\hat P^{(a)}_i,\hat P^{(b)}_j\big)
\quad\text{with}\quad
\hat P^{(\cdot)}_i=\frac{1}{|\mathbf{F}_i|}\sum_{\mathbf{f}\in \mathbf{F}_i}\delta_{\mathbf{f}}
\end{equation}
where $\mathsf{SW}_\varepsilon$ is the entropically-regularized $2$-Wasserstein distance computed under a squared Euclidean ground cost in feature space. We employ a stabilized Sinkhorn solver with median-scaled costs and a fixed stopping tolerance, ensuring numerical stability across clusters of varying size.
This construction naturally handles single-vertex clusters and multi-vertex clusters in a unified way.
Beyond feature similarity, we further encode the relative geometric configuration of the 3HG clusters on $\mathbb{S}^2$.
Let $\mathbf{U}^{3HG}_i=\{\mathbf{u}_p\}_{p\in \Omega_i^{(k)}}\subset\mathbb{S}^2$ denote the unit vectors of the 3HGs within cluster $i$. 
The cluster-level geometric discrepancy is quantified using a spherical Chamfer distance computed on great-circle metrics:
\begin{equation}
D_{\mathrm{geo}}(i,j)
=
\frac{1}{\lvert \mathbf{U}^{\mathrm{3HG}}_i \rvert}
\sum_{\mathbf{u}\in \mathbf{U}^{\mathrm{3HG}}_i}
\min_{\mathbf{v}\in \mathbf{U}^{\mathrm{3HG}}_j}
d_{\mathbb{S}^2}(\mathbf{u},\mathbf{v})
\;+\;
\frac{1}{\lvert \mathbf{U}^{\mathrm{3HG}}_j \rvert}
\sum_{\mathbf{v}\in \mathbf{U}^{\mathrm{3HG}}_j}
\min_{\mathbf{u}\in \mathbf{U}^{\mathrm{3HG}}_i}
d_{\mathbb{S}^2}(\mathbf{u},\mathbf{v}) \, 
\end{equation}
where $d_{\mathbb{S}^2}$ is the great-circle distance.
We optionally apply a centroidal geodesic threshold $\theta_g$ to gate implausible pairs, assigning a large penalty to cluster pairs with centroid distances exceeding $\theta_g$.

To mitigate mismatches in densely folded cortical regions, two cluster-level regularizers are introduced based on the spherical centroids (mean of $\mathbf{U}_i$ on $\mathbb{S}^2$): a local density mismatch and a $k$-NN context mismatch.
Let $\rho_i$ denote a Gaussian-kernel density estimate computed from the $k_{\mathrm{den}}$ nearest centroids with bandwidth $\sigma_{\mathrm{den}}$ in degrees,
and let $\boldsymbol{\nu}_i \in \mathbb{R}^{k_{\mathrm{ctx}}}$ represent the sorted vector of geodesic distances to the $k_{\mathrm{ctx}}$ nearest centroids.
We define
\begin{equation}
D_{\mathrm{dens}}(i,j)=|\rho_i-\rho_j|,
\qquad
D_{\mathrm{ctx}}(i,j)=\big\|\boldsymbol{\nu}_i-\boldsymbol{\nu}_j\big\|_2
\end{equation}


For two subjects $a$ and $b$, the total correspondence cost integrates the morphological, geometric, and contextual components:
\begin{equation}
C_{ij}\;=\;\lambda_{geo}\,D_{\mathrm{geo}}(i,j)\;+\;\lambda_{feat}\,D_{\mathrm{feat}}(i,j)\;+\;\lambda_{dens}\,D_{\mathrm{dens}}(i,j)\;+\;\lambda_{ctx}\,D_{\mathrm{ctx}}(i,j)
\label{eq::correspondence}
\end{equation}
with nonnegative weights $\lambda_{geo},\lambda_{feat},\lambda_{dens},\lambda_{ctx}$.
We then solve the linear assignment problem:
\begin{equation}
\min_{P\in\{0,1\}^{K\times K}}\ \langle C,P\rangle
\quad\text{s.t.}\quad
P\mathbf{1}=\mathbf{1},\ \ P^\top \mathbf{1}=\mathbf{1}
\end{equation}
via the Hungarian algorithm to obtain a bijection $\pi$ between the $K$ clusters:
$P_{i\,\pi(i)}=1$.
This yields a cross-subject one-to-one correspondence consistent with both local morphology and spherical geometry, while being regularized by regional crowding and neighborhood context.
For a cohort, we fix one subject as the template and compute the above assignment between the anchor and each target subject independently, thereby defining a globally consistent label space without pairwise rematching.
This anchor-based strategy offers several practical advantages. First, it reduces computational complexity, rather than performing pair-wise alignments for every subject pair in the cohort, we only compute mappings between the anchor and each other subject. Second, it guarantees the global consistency of cluster labels. Once the labels of the anchor subject are set, no further relabelling is required when new subjects are added. Finally, this approach supports incremental expansion of the cohort. Whenever a new individual enters the study, we need only compute a single mapping between that subject and the anchor, leaving all existing correspondences untouched. This design both simplifies longitudinal or multi-site studies and preserves the integrity of previously established cluster relationships.
\section{Experiments and Results}
\subsection{Dataset Setup}
To evaluate our framework, we used T1-weighted structural MRI (sMRI) and diffusion tensor imaging (DTI) data from 1064 subjects in the Human Connectome Project S1200 release \cite{VANESSEN201362}. 
Data preprocessing of sMRI data was performed using FMRIB Software Library (FSL)~\cite{JENKINSON2012782} and FreeSurfer~\cite{FISCHL2012774}. 
In brief, We first resampled the three-dimensional T1-weighted volumes to 1 mm isotropic resolution using cubic spline interpolation. 
Then, skull removal, gray matter (GM)/white matter (WM) tissue segmentation, and WM surface reconstruction were performed. During surface reconstruction, the interface between GM and WM was modeled as a triangle mesh in 3D space. 
Each subject’s preprocessed cortical surfaces were onto the spherical
space and further resampled to a standard cortical surface~\cite{FISCHL2012774}, corresponding to the 7th subdivision (fsaverage7) of an icosahedron with 163,842 vertices per hemisphere, following previous methods~\cite{zhao2019spherical,zhao2021spherical,hu2024consecutive}.
We then applied the computational framework described in \cite{chen2017gyral,cao2026gyral} to extract GyralNet, the whole-brain representation of gyral folding organization, and subsequently identified the 3HGs. The statistical properties of the resulting GyralNet are summarized in Table \ref{tab:hemisphere_stats}.
For cortical parcellation and ROI labeling of the 3HGs, we employed the Desikan-Killiany (DK) atlas \cite{desikan2006automated} and the Destrieux atlas \cite{destrieux2010automatic,fischl2004automatically} supplied by FreeSurfer.
DTI data were corrected for eddy currents using FSL and fiber tracking was performed with DSI Studio \cite{yeh2013deterministic}.
After registering the sMRI data to DTI space,
the trace-map descriptor and the structural similarity of each 3HG on the reconstructed white surface were then calculated.

\begin{table*}[htbp]
\centering
\footnotesize
\setlength{\tabcolsep}{1pt}
\renewcommand{\arraystretch}{1.15}
\caption{Summary statistics of GyralNet graphs per hemisphere on 1,056 HCP subjects.}
\label{tab:hemisphere_stats}
\begin{tabular}{lccccc}
\toprule
\multirow{2}{*}{\textbf{Metric}} &
\multicolumn{2}{c}{\textbf{Left Hemisphere}} &
\multicolumn{2}{c}{\textbf{Right Hemisphere}} &
\multirow{2}{*}{\textbf{Combined mean $\pm$ SD}} \\
\cmidrule(lr){2-3} \cmidrule(lr){4-5}
& Range & Mean $\pm$ SD & Range & Mean $\pm$ SD & \\
\midrule
{Node count} & [108, 222] & 158.97 $\pm$ 18.48 & [112, 233] & 158.96 $\pm$ 18.21 & 158.96 $\pm$ 18.35 \\
{Edge count} & [149, 315] & 224.35 $\pm$ 18.71 & [153, 335] & 224.80 $\pm$ 18.64 & 224.57 $\pm$ 18.68 \\
{Deg-1 nodes} & [0, 7] & 0.95 $\pm$ 0.96 & [0, 5] & 0.85 $\pm$ 0.88 & 0.90 $\pm$ 0.92 \\
{Deg-2 nodes} & [10, 47] & 26.32 $\pm$ 5.86 & [11, 43] & 25.61 $\pm$ 5.73 & 25.96 $\pm$ 5.81 \\
{Deg-3 nodes} & [83, 185] & 131.50 $\pm$ 16.92 & [84, 204] & 132.31 $\pm$ 16.70 & 131.91 $\pm$ 16.82 \\
\addlinespace[0.3em]
{Total nodes} & \multicolumn{2}{c}{167,872} & \multicolumn{2}{c}{167,860} & 335,732 \\
{Total edges} & \multicolumn{2}{c}{236,916} & \multicolumn{2}{c}{237,384} & 474,300 \\
\bottomrule
\end{tabular}
\end{table*}

\subsection{Implementation Details}
To compute the trace map descriptor for each 3HG, we first extract all fiber bundles passing through the corresponding vertex. Each fiber is then divided into overlapping segments by applying a sliding window of length 8 with a step size of 4. For each segment, we estimate its principal direction and project that direction vector onto the unit sphere.
And the trace-map descriptor of the point is obtained by counting the proportion of the direction vectors in the 144 regions evenly divided on the sphere.
In~\cite{chen2025using}, the introduced structural similarity of each 3HG is represented as a hierarchical multi-hop feature set that encodes its topological profile. Since only a naive topological summary is required in this work and high-dimensional hierarchies are unnecessary, we restrict this to one-hop structural similarity, thereby reducing feature complexity. The dimension of this topological profile matches the number of regions in the atlas used for ROI labeling of the 3HGs. 
Consequently, for $n$ 3HGs on a cortical hemisphere, the combined dual profile feature matrix $\textbf{X}$ has dimensions $ n\times(144+b)$, where $b$ is the number of brain regions of a given atlas on a hemisphere.
The graph convolutional network models were trained from scratch using two graph convolutional layers separated by a 256-dimensional intermediate representation and SeLU~\cite{klambauer2017self} as the activate function. The dimension of the node embedding, $r$, was set equal to the number of atlas hemispheres, \textit{i.e.} $b$. 
Training proceeded for up to 1,500 iterations per hemisphere, with early termination applied once the standard deviation of the loss over the most recent 20 iterations fell below $1\times{10}^{-3}$. 
During training, the hyperparameters $\alpha$, $\beta$ and $\gamma$ in Eq.~\ref{eq:loss} were set at 1, 0.1 and 0.1, respectively. 
For JMGM, the weight $\lambda_{geo},\lambda_{feat},\lambda_{dens},\lambda_{ctx}$ in Eq.~\ref{eq::correspondence} were set at 0.3, 0.7, 0.2 and 0.2 after exhaustive grid search.
All models were implemented in PyTorch and the experiments were conducted on a workstation equipped with an NVIDIA RTX 6000 Ada GPU and a 3.6 GHz Intel processor.
All experimental results were obtained using a setup with 48 clusters ($k=48$). Experiments regarding the choice of $k$ can be found in Section~\ref{sec::bestk}.
\subsection{Evaluation on the Intra-Subject 3HG Clustering Performance}
\label{sec::clustering}
\begin{table*}[t]
\centering
\begin{threeparttable}
\caption{Comparison of clustering performance between the proposed framework and baseline methods. The clustering features include topological context ($\mathcal{T}$) and structural fiber connectivity patterns ($\mathcal{C}$), evaluated both jointly and separately. Performance is quantified using four metrics: global conductance (COND), coverage (COV), diffusion rate (DR), and modularity (MOD). }
\label{tab:clustering}

\begingroup
\setlength{\tabcolsep}{2.5pt}   
\renewcommand{\arraystretch}{1.1}
\scriptsize                     

\begin{tabular*}{\textwidth}{@{\extracolsep{\fill}}%
l l
c c c c 
c c c c  
@{}}
\toprule
\multicolumn{2}{c}{\textbf{Methods}} &
\multicolumn{4}{c}{\textbf{Destrieux Atlas}} &
\multicolumn{4}{c}{\textbf{Desikan-Killiany Atlas}} \\
\cmidrule(lr){1-2}\cmidrule(lr){3-6}\cmidrule(lr){7-10}
\textbf{Feature} & \textbf{Model} &
\textbf{COND}$\downarrow$\ & \textbf{COV}$\uparrow$ & \textbf{DR}$\downarrow$ & \textbf{MOD}$\uparrow$  &
\textbf{COND}$\downarrow$ & \textbf{COV}$\uparrow$ & \textbf{DR}$\downarrow$ & \textbf{MOD}$\uparrow$  \\
\midrule
\multirow{4}{*}{$\mathcal{T} + \mathcal{C}$}
  & Ours     & \underline{0.581} & \textbf{0.572} & \underline{24.93} & \textbf{0.543}       & \textbf{0.522} & \textbf{0.683} & \textbf{25.50} & 0.521    \\
  & DiffPool   & 0.673 & 0.524 & 26.22 & \underline{0.540}  & 0.670 & 0.562 & \underline{25.78} & \textbf{0.546}\\
 & DeepWalk   & \textbf{0.573} & 0.563 & 29.47 & 0.487 & \underline{0.531} & \underline{0.605} & 27.83 & 0.482\\
  & DMoN   & 0.591 & \underline{0.568} & \textbf{22.28} & 0.539  & 0.558 & 0.551 & 26.55 & 0.501 \\
\addlinespace[2pt]
\hline
\multirow{4}{*}{$\mathcal{T}$ only}
  & Ours     & 0.582 & 0.483 & 27.33 & 0.445   & 0.679 & 0.581 & 26.54 & 0.487    \\
  & DiffPool  & 0.609 & 0.531 & 28.06 & 0.470  & 0.651 & 0.559 & 26.38 & \underline{0.542}   \\
& DeepWalk   & 0.599 & 0.541 & 25.89 & 0.481 & 0.631 & 0.562 & 26.13 & 0.498   \\
& DMoN   & 0.590 & 0.539 & 26.01 & 0.500  & 0.551 & 0.590 & 27.11 & 0.516   \\
\addlinespace[2pt]
\hline
\multirow{4}{*}{$\mathcal{C}$ only}
  & Ours     & 0.655 & 0.510 & 25.25 & 0.498  & 0.655 & 0.510 & 26.25 & 0.498 \\
  & DiffPool   & 0.632 & 0.533 & 26.11 & 0.527 & 0.632 & 0.533 & 26.11 & 0.527\\
  & DeepWalk  & 0.604 & 0.527 & 28.69 & 0.522  & 0.604 & 0.527 & 28.69 & 0.522   \\
  & DMoN   & 0.598 & 0.506 & 25.33 & 0.519  & 0.598 & 0.506 & 26.33 & 0.519   \\
\bottomrule
\end{tabular*}
\endgroup

\end{threeparttable}
\end{table*}
To evaluate the clustering performance of the proposed framework, we compare our method with three baselines:
\begin{itemize} 
\item \textbf{DeepWalk}~\cite{perozzi2014deepwalk} represents a naïve strategy of collecting node features to learned node embeddings of the graph. clustering is then performed using the classical k‐means algorithm via the Lloyd procedure~\cite{lloyd1982least} with k-means++ initialization~\cite{arthur2006k}.
\item  \textbf{DMoN}~\cite{tsitsulin2023graph} is a recently proposed unsupervised GNN-based clustering method that optimizes a differentiable modularity‐inspired objective to derive soft cluster assignments for attributed graphs.
\item \textbf{DiffPool}~\cite{ying2018hierarchical} is a hierarchical graph‐pooling method that learns differentiable soft‐assignment matrices to group nodes into clusters across layers, thereby producing coarsened graph representations in a GNN framework.
\end{itemize}
In addition, to validate the effectiveness of the proposed dual-profile feature representation, we report the results of solely employing the topological context ($\mathcal{T}$) and structural fiber connectivity pattern ($\mathcal{C}$) respectively.
To evaluate the clustering performance, four complementary network-based measures are employed: the global conductance (COND), coverage (COV), diffusion rate (DR), and modularity (MOD). These measures jointly characterize the balance between network integration and segregation within the GyralNet.
The COND metric quantifies the efficiency of information flow within clusters by measuring how well vertices sharing similar gyral-folding features are integrated within each cluster and how isolated each cluster is from the rest of the network. A lower COND value indicates stronger intra-cluster homogeneity and higher morphological coherence, reflecting improved local integration of vertices sharing similar gyral folding features.

As shown in Table~\ref{tab:clustering}, we report results using anatomical location encoding based on DK atlas and Destrieux atlas as topological features for clustering. It's important to note that because atlas information is only used within the topological context, the results from both atlases will be the same when only the structural connectivity profile ($\mathcal{C}$ only) is used.
Compared to the baseline models, our method shows lower COND and DR values and higher COV scores across both the Destrieux and DK atlases. The results obtained using the combined features ($\mathcal{T}+\mathcal{C}$) generally exhibit the best overall performance among all settings. For the Destrieux atlas, our method achieves the lowest COND and DR, together with the highest COV and MOD, while similar trends are observed in the DK atlas. When using only the topological feature or connectivity feature the performance of all methods shows a moderate decrease compared with the combined-feature setting.

\begin{figure}[h!]
    \centering
    \begin{minipage}[t]{0.49\textwidth} 
        \centering
        \includegraphics[width=\textwidth]{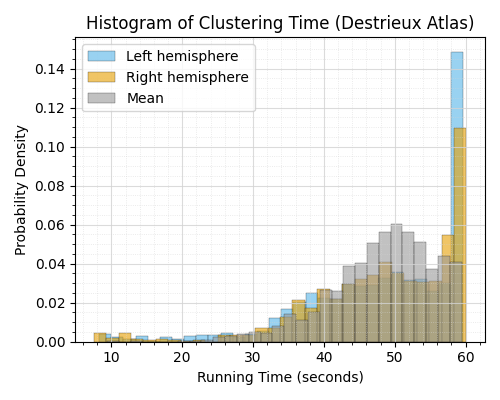} 
        \label{fig:left}
    \end{minipage}%
    \hfill 
    \begin{minipage}[t]{0.49\textwidth} 
        \centering
        \includegraphics[width=\textwidth]{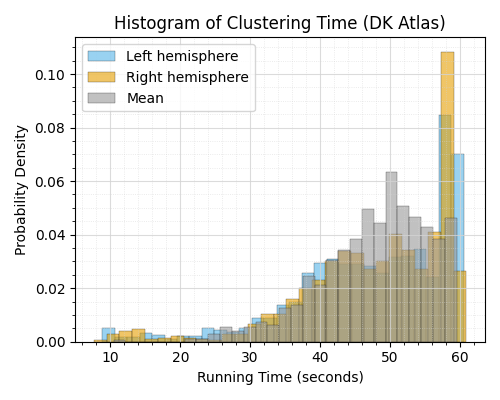} 
        \label{fig:right}
    \end{minipage}
            \caption{Histograms of clustering time for the proposed method using the Destrieux (left) and DK (right) atlases, shown separately for the left hemisphere, right hemisphere, and their mean.}
\label{fig::cluster_time}
\end{figure}
\begin{figure*}[h!]
    \centering

    \begin{subfigure}[c]{0.48\textwidth}
        \centering
        \includegraphics[width=\linewidth]{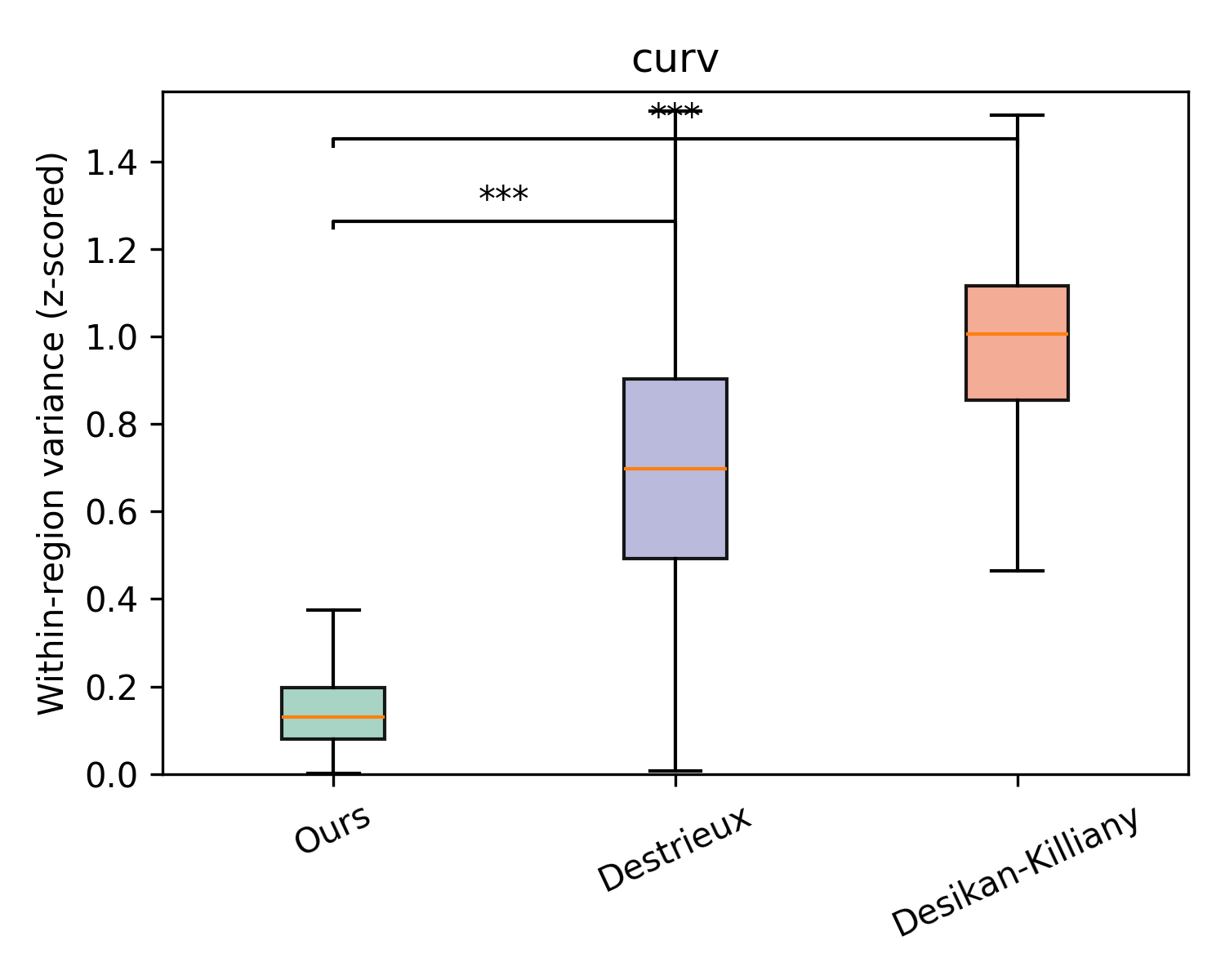}
    \end{subfigure}
    \hfill
    \begin{subfigure}[c]{0.48\textwidth}
        \centering
        \includegraphics[width=\linewidth]{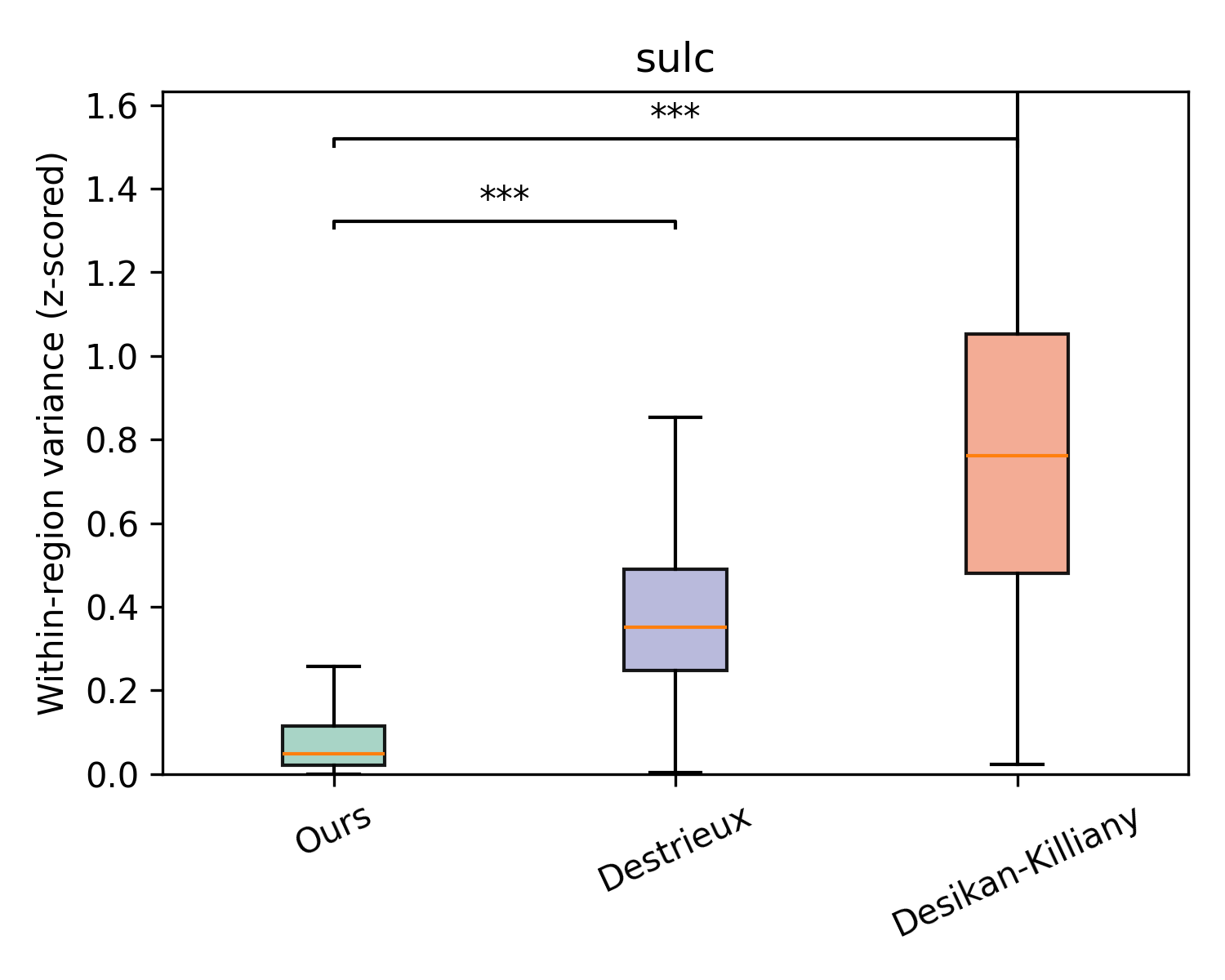}
    \end{subfigure}

    \begin{subfigure}[c]{0.48\textwidth}
        \centering
        \includegraphics[width=\linewidth]{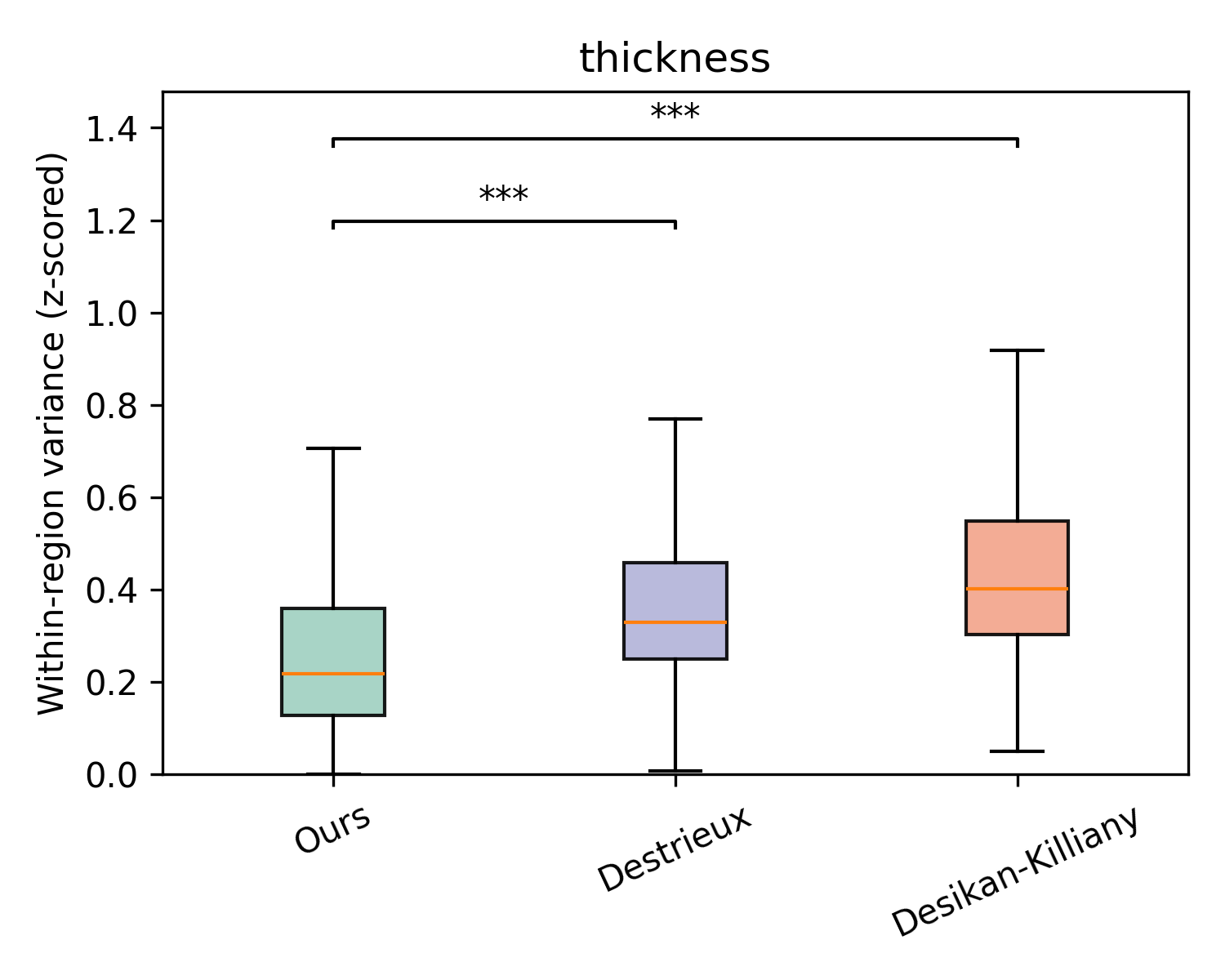}
    \end{subfigure}
    \hfill
    \begin{subfigure}[c]{0.48\textwidth}
        \centering
        \includegraphics[width=\linewidth]{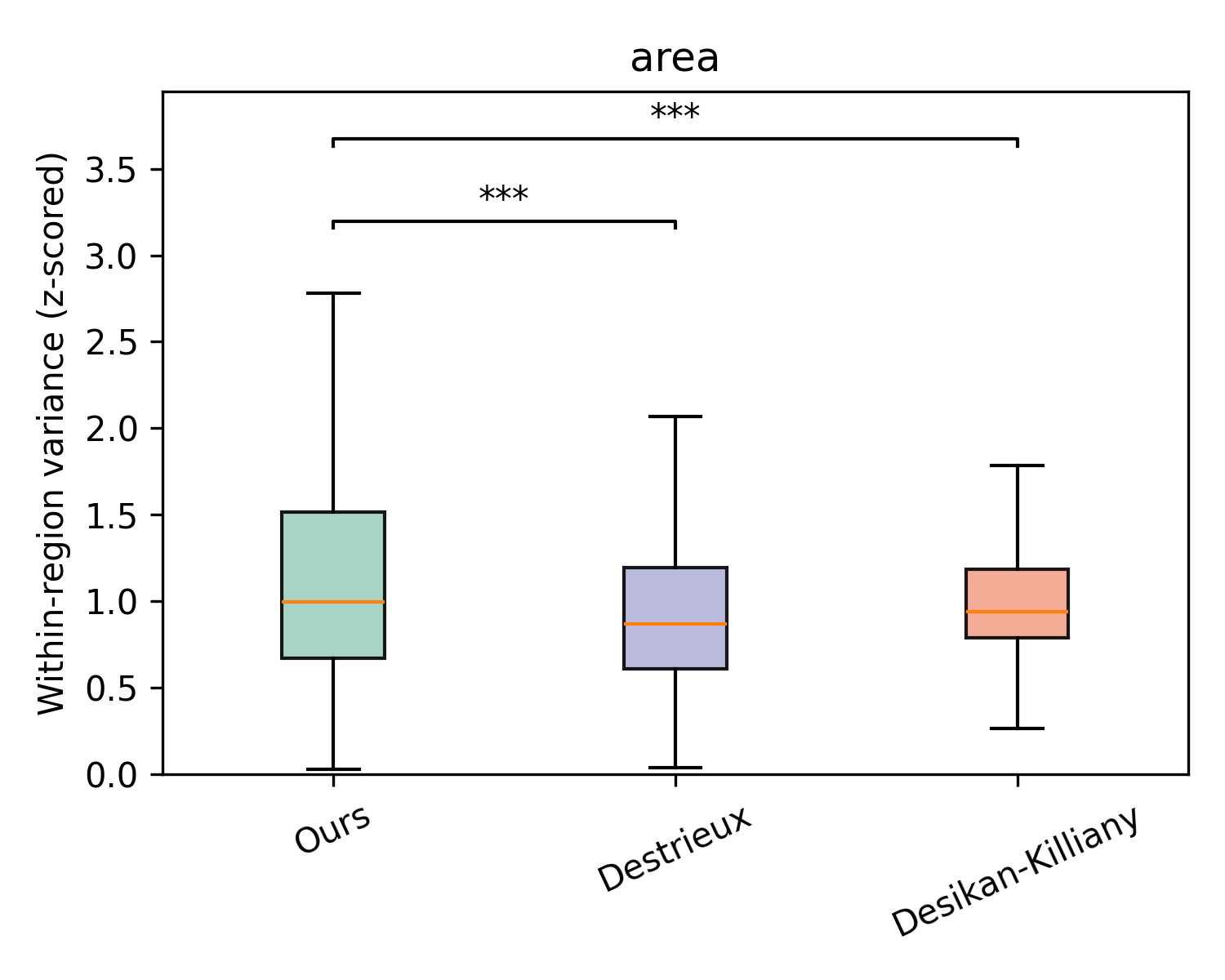}
    \end{subfigure}
\vspace{-0.5cm}
    \caption{Boxplots show the within region variance of cortical features for the proposed GyralNet clusters, the Destrieux atlas and the DK atlas after z-score normalization. GyralNet subnetworks consistently display lower variance across curvature, sulcal depth, thickness and area. This indicates that the proposed framework produces more homogeneous regions compared with traditional anatomical atlases. Significance marks ($^{***}$) indicate statistical differences ($p<0.001$) based on the Mann-Whitney U-test.}
    \label{fig:homo}
\end{figure*}

In Fig.~\ref{fig::cluster_time}, we plot the histograms of clustering time for the proposed method using the Destrieux and DK atlases, shown separately for the left and right hemispheres as well as their mean. The distributions show similar patterns across atlases, with most clustering runs concentrated between 40 s and 60 s.
To evaluate the anatomical coherence of the 3HG clusters, we compared their within-region variance of classical cortical morphometry against atlas-based ROI segmentation.
The result is shown in Fig~\ref{fig:homo}, across all four features, subnetworks consistently showed significantly lower morphometric variance than ROIs, demonstrating that our segmentation produces more homogeneous and anatomically meaningful regions.

\subsection{Evaluation on the Cross-Subject Correspondence of 3HG Clusters}
\label{sec::correspondence}
\begin{figure*}[htb]

  \centering
  \centerline{\includegraphics[width=\linewidth]{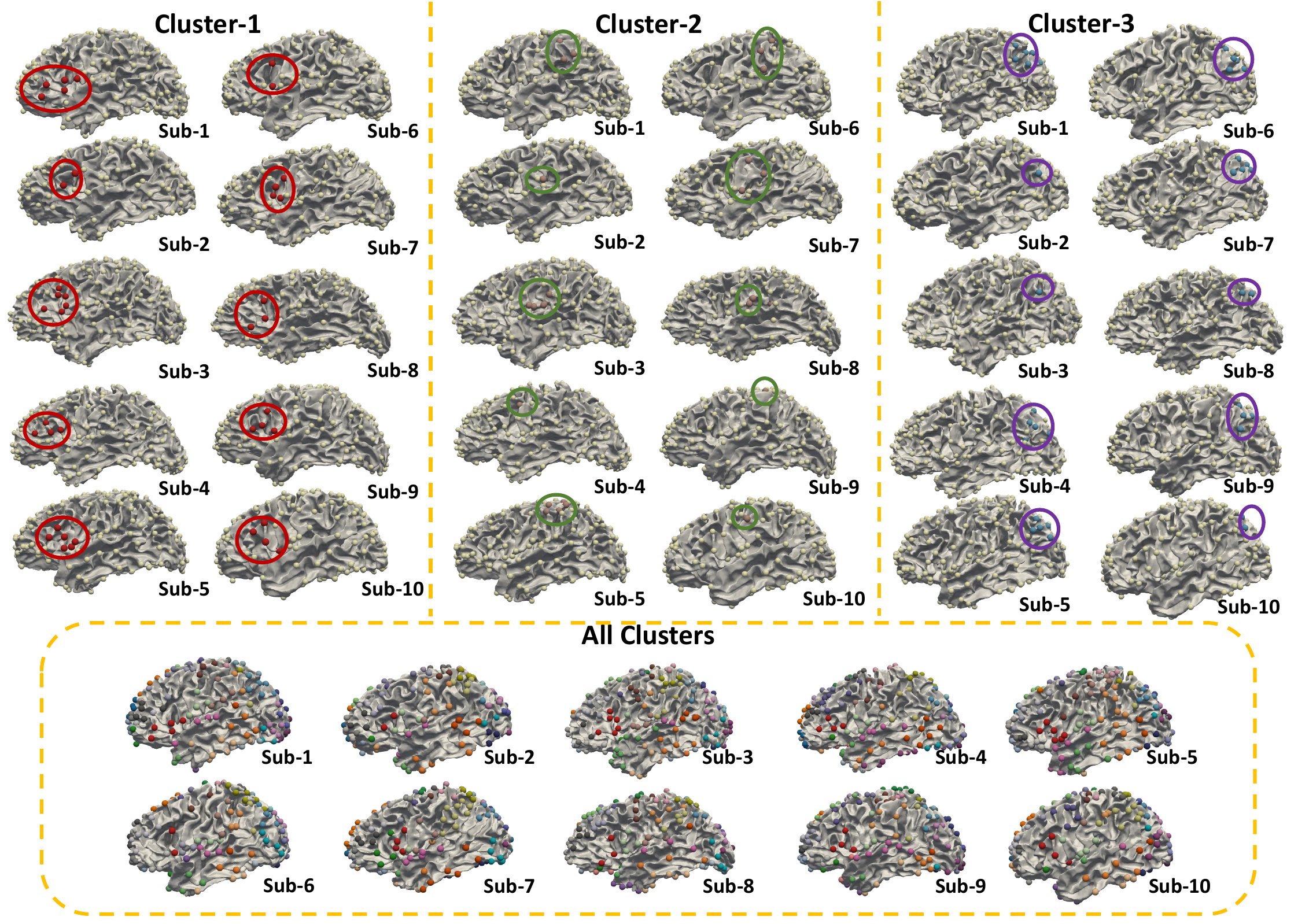}}
\vspace{-0.3cm}
\caption{Qualitative visualization of cross-subject correspondences established by the proposed JMGM using the Destrieux atlas. Three representative clusters (Cluster-1, Cluster-2, and Cluster-3) are shown across ten subjects, where consistent gyral patterns are highlighted in red, green, and purple, respectively. The bottom panel displays the full set of identified 3HG clusters for each subject, demonstrating stable and anatomically coherent correspondences across individuals.}
\label{fig:correspondence_destrieux}
\end{figure*}
In this section, we evaluate the performance of the proposed JMGM strategy in establishing cross-subject correspondence of 3HGs. Following the evaluation protocol in previous work~\cite{chen2025using}, we report the ROI hit rate as a measure of alignment accuracy. 
A gyral-sulcal-based atlas~\cite{destrieux2010automatic} is adopted for ROIs parcellation.
We compare the proposed JMGM with three previously proposed methods for cortical folding correspondence developed by our group, as well as one classical registration-based method, Iterative Closest Point (ICP)~\cite{besl1992method}, which is among the most widely used algorithms for point cloud alignment:
\begin{itemize}
    \item \textbf{Cortex2vec}~\cite{zhang2023cortex2vector} is a learning-based embedding framework that encodes cortical folding patterns into anatomically meaningful latent vectors. 
It captures both individual variability and population-level regularities by mapping local cortical landmarks into a shared representation space.
    For cross-subject correspondence, each cluster is represented by the mean embedding of its constituent 3HGs, and inter-cluster similarity is computed using cosine similarity between these averaged representations.
    \item 
    \textbf{CortexSSE}~\cite{chen2025using} is an extension of Cortex2vec that improves the representation of local topological structures and enhances embedding robustness through a selective reconstruction loss. The correspondence construction process follows the same procedure as Cortex2vec, differing only in the use of structurally enhanced embeddings.
    \item 
    \textbf{GyralParti}~\cite{zhuang2025gyralnet} uses a localization descriptor to represent each subnetwork of 3HGs within the GyralNet, and inter-cluster correspondence across subjects is established via cosine similarity followed by Hungarian assignment.
    \item \textbf{ICP}~\cite{besl1992method} serves as a geometry-based baseline that establishes correspondences through iterative rigid alignment on the inflated spherical surface. Each 3HG cluster is represented by its centroid on $\mathbb{S}^2$, and correspondences are computed by minimizing geodesic distances between matched centroids. The transformation is refined through iterative updates until convergence, yielding purely geometry-driven mappings without incorporating morphological or contextual information.
    
\end{itemize}

\begin{table*}[t]
\centering
\begin{threeparttable}
\caption{ROI hit rate (mean$\pm$std) of cross-subject correspondence for different clustering methods using one-hot feature embeddings derived from the Destrieux and Desikan–Killiany atlases. Results are reported separately for the left and right hemispheres and their averages. The proposed JMGM achieves the highest correspondence accuracy across both feature sources. }
\label{tab:correspondence}

\begingroup
\setlength{\tabcolsep}{2.5pt}   
\renewcommand{\arraystretch}{1.1}
\scriptsize                     

\begin{tabular*}{\textwidth}{@{\extracolsep{\fill}}%
l c
c c c  
c c c 
@{}}
\toprule
\multicolumn{1}{c}{\multirow{2}{*}{\textbf{Methods}}}&
\multicolumn{3}{c}{\textbf{Destrieux Atlas}} &
\multicolumn{3}{c}{\textbf{Desikan-Killiany Atlas}} \\
\cmidrule(lr){2-4}\cmidrule(lr){5-7}
&  
\textbf{L. Hemi.} & \textbf{R. Hemi.}  & \textbf{Average}   &
\textbf{L. Hemi.} & \textbf{R. Hemi.}  & \textbf{Average}  \\
\midrule
\textbf{Cortex2vec}~\cite{zhang2023cortex2vector}   & 0.546$\pm$0.02 & 0.572$\pm$0.03 & 0.559$\pm$0.02 & 0.576$\pm$0.04      & 0.590$\pm$0.03 & 0.584$\pm$0.04   \\
\textbf{CortexSSE}~\cite{chen2025using}   & 0.603$\pm$0.03 & 0.579$\pm$0.05 & 0.590$\pm$0.04 & 0.595$\pm$0.03  & 0.612$\pm$0.03 & 0.604$\pm$0.03 \\
\textbf{GyralParti}~\cite{zhuang2025gyralnet}   & 0.747$\pm$0.06 & 0.687$\pm$0.07 & 0.717$\pm$0.04 & 0.781$\pm$0.04  & 0.756$\pm$0.04 & 0.769$\pm$0.03 \\
\textbf{ICP}~\cite{besl1992method}   & 0.735$\pm$0.06 & 0.687$\pm$0.07 & 0.711$\pm$0.04 & 0.770$\pm$0.07  & 0.817$\pm$0.06 & 0.793$\pm$0.05 \\
\textbf{JMGM } (Ours)  & \textbf{0.807$\pm$0.04} & \textbf{0.797$\pm$0.04}& \textbf{0.802$\pm$0.03} & \textbf{0.789$\pm$0.07}  & \textbf{0.841$\pm$0.06} & \textbf{0.815$\pm$0.04} \\
\bottomrule
\end{tabular*}
\endgroup

\end{threeparttable}
\end{table*}

As shown in Table~\ref{tab:correspondence}, we randomly select 10 subjects as the anchor subject for constructing correspondence, and report the average ROI hit rate of cross-subject correspondence for different clustering methods using one-hot feature embeddings derived from the Destrieux and DK atlases.
The proposed JMGM consistently achieves the highest ROI hit rates across both the Destrieux and DK atlases. In the Destrieux atlas, JMGM reaches $0.807 \pm 0.04$ for the left hemisphere, $0.797 \pm 0.04$ for the right hemisphere, and $0.802 \pm 0.03$ on average, outperforming all baseline methods. A similar trend is observed for the Desikan–Killiany atlas, where JMGM attains $0.789 \pm 0.07$, $0.841 \pm 0.06$, and $0.815 \pm 0.04$, respectively. Among the baselines, GyralParti and ICP show relatively high accuracy, while Cortex2Vec and CortexSSE perform lower overall. The consistent improvements across both atlases indicate that JMGM provides more reliable and anatomically consistent cross-subject correspondences.
Figure~\ref{fig:correspondence_destrieux} shows qualitative results of cross-subject correspondences obtained using the Destrieux atlas. The visualized clusters exhibit consistent spatial localization across subjects, indicating that JMGM preserves anatomically coherent 3HG patterns at the group level.

\subsection{Ablation Studies}
\subsubsection{Hyperparameter Sensitivity}
\begin{figure*}[h!]
    \centering

    \begin{subfigure}[c]{0.48\textwidth}
        \centering
        \includegraphics[width=\linewidth]{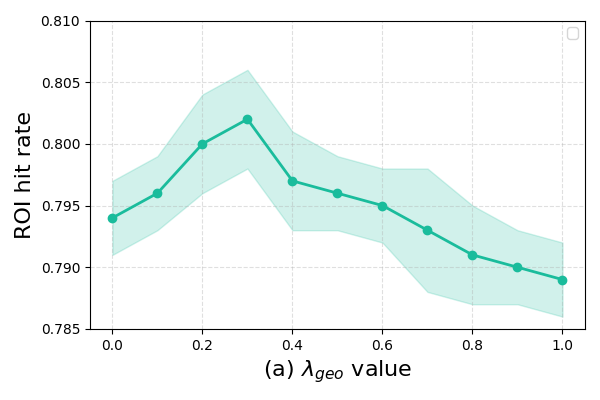}
        \label{fig:a}
    \end{subfigure}
    \hfill
    \begin{subfigure}[c]{0.48\textwidth}
        \centering
        \includegraphics[width=\linewidth]{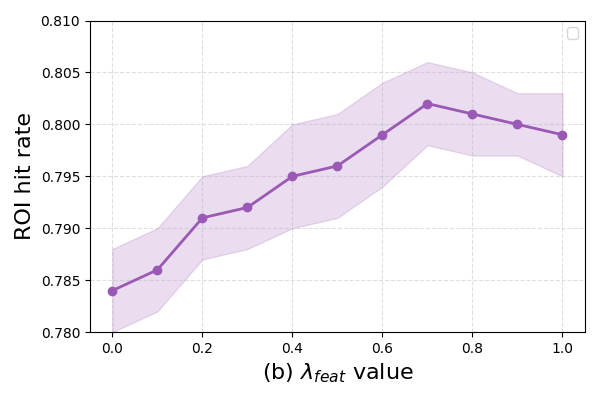}
        \label{fig:b}
    \end{subfigure}

    \begin{subfigure}[c]{0.48\textwidth}
        \centering
        \includegraphics[width=\linewidth]{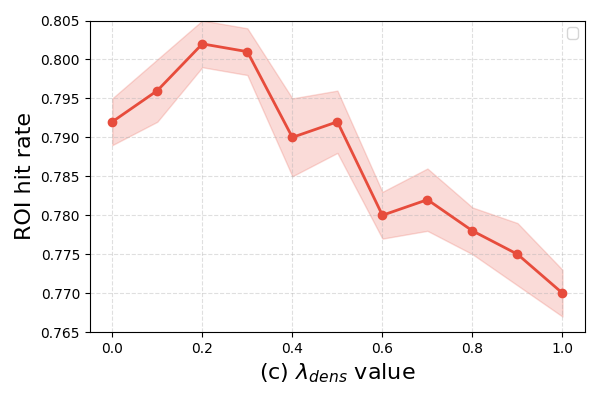}
        \label{fig:c}
    \end{subfigure}
    \hfill
    \begin{subfigure}[c]{0.48\textwidth}
        \centering
        \includegraphics[width=\linewidth]{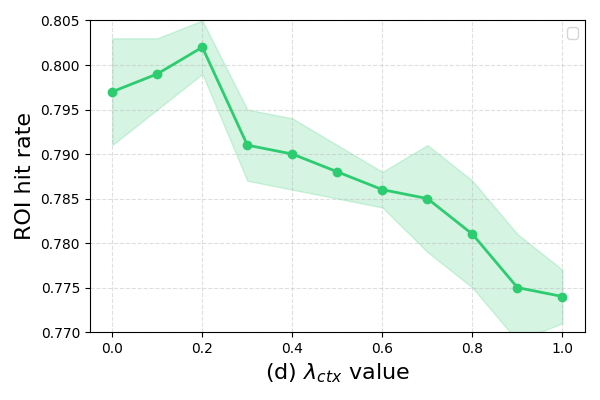}
        \label{fig:d}
    \end{subfigure}

    \caption{Sensitivity analysis of the four weighting coefficients in the correspondence cost. Each panel reports the ROI hit rate as a function of a single parameter while keeping the remaining coefficients fixed: (a) geometric discrepancy weight 
$\lambda_{geo}$, (b) feature similarity weight $\lambda_{feat}$, (c) local density mismatch weight $\lambda_{dens}$, and (d) $k$-NN context mismatch weight $\lambda_{ctx}$. Shaded regions denote the standard deviation across subjects.}
    \label{fig:4panels}
\end{figure*}

Figure~\ref{fig:4panels} reports the ROI hit rate as each weighting coefficient in the correspondence cost is varied independently. 
For  $\lambda_{geo}$ and $\lambda_{feat}$, the hit rate increases initially and then stabilizes or slightly decreases at higher values. In contrast, varying $\lambda_{dens}$ and $\lambda_{ctx}$ produce a more monotonic decline after the peak at lower parameter values. Across all four panels, the curves exhibit smooth changes with relatively narrow variability bands, indicating consistent behavior across subjects.

\subsubsection{Ablation Analysis of the Clustering Loss Components and Topological Refinement}
\begin{table*}[t]
\footnotesize
\centering
\caption{Ablation analysis of the clustering loss components and topological refinement, reported using within-region variance (mean $\pm$ std).}
\label{tab:variance_summary}
\begin{tabular}{lcccc}
\toprule
 & \textbf{Curv} & \textbf{Sulc} & \textbf{Thickness} & \textbf{Area} \\
\midrule
w/o $\mathcal{L}_p$     
& 0.171 $\pm$ 1.307 & 0.095 $\pm$ 0.093& 0.473 $\pm$ 0.195 & 1.207 $\pm$ 0.801\\
w/o $\mathcal{L}_{det}$     
& 0.175 $\pm$ 1.321 & 0.092 $\pm$ 0.099& 0.462 $\pm$ 0.246 & 1.201 $\pm$ 0.793 \\
w/o $\mathcal{L}_{c}$     
& 0.167 $\pm$ 1.179 & 0.089 $\pm$ 0.101 & 0.457 $\pm$ 0.208 & 1.199 $\pm$ 0.684 \\
w/o refinement    
& 0.176 $\pm$ 1.305 & 0.094 $\pm$ 0.103&\textbf{0.276 $\pm$ 0.301} & \textbf{1.066 $\pm$ 0.747} \\
\hline
Ours   
& \textbf{0.161 $\pm$ 1.299} & \textbf{0.088 $\pm$ 0.104} & 0.452 $\pm$ 0.215 & 1.196 $\pm$ 0.780 \\
\bottomrule
\end{tabular}
\end{table*}
Table~\ref{tab:variance_summary} summarizes the ablation analysis conducted to evaluate the contribution of individual loss components and the topological refinement step. The assessment is based on within-region variance computed across four morphometric features. Removing any single loss term leads to increased variance for most features, indicating reduced intra-cluster homogeneity. For curvature and sulcal depth, the full model yields the lowest variance among all configurations. For thickness and area, the differences across settings are smaller, and the variant without refinement produces slightly lower variance than the full model. Overall, the complete formulation achieves consistently competitive or best performance across features, demonstrating that each loss component contributes to stable cluster formation.
Note that curvature exhibits higher across-region variability than other surface features, which is expected given its inherently high-frequency nature. Our z-score normalization is applied at the vertex level within each subject; therefore, absolute variance magnitudes differ across features, while comparisons among methods remain valid.
\subsubsection{Evaluating and Choosing the Best $k$ for Clustering}
\label{sec::bestk}
Determining the optimal number of clusters, $k$, is a critical step for ensuring that the resulting cortical folding patterns correspondence are both anatomically meaningful and consistent across subjects. 
In this study, we systematically evaluate different candidate values of $k$ to balance the trade-off between over-partition and under-partition of gyral folding patterns. 
To quantitatively assess the anatomical coherence and topological stability of the clustering results,
we evaluate all methods using the four clustering metrics described in Sec.~\ref{sec::clustering} across candidate numbers of clusters (
$k=16–72$).
\begin{figure}[htb]

  \centering
  \centerline{\includegraphics[width=\linewidth]{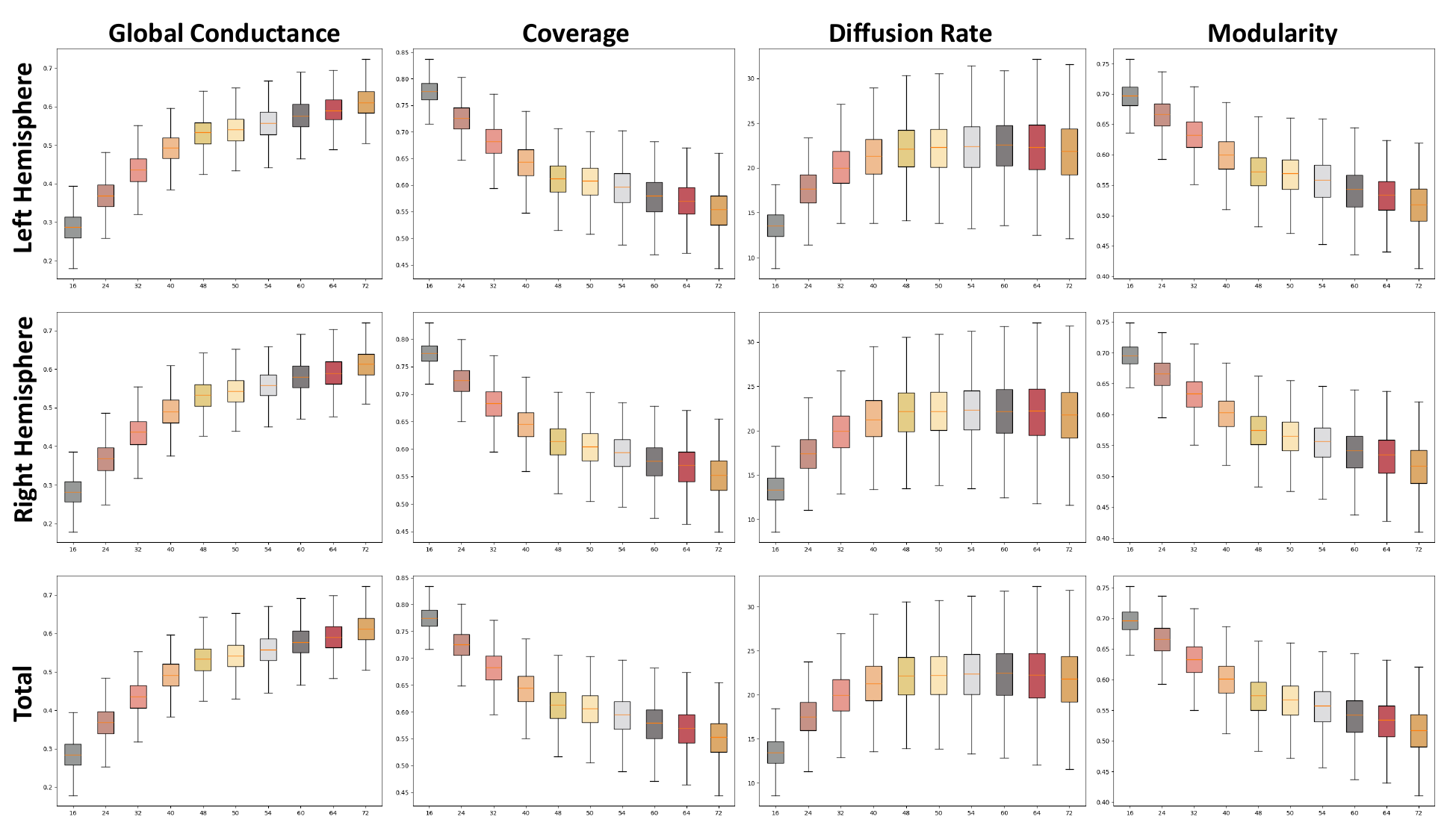}}
\vspace{-0.3cm}
\caption{Distributions of four network measures across different cluster numbers $k$ for both hemispheres.}
\label{fig:cluster}
\end{figure}
As shown in Fig.~\ref{fig:cluster}, these four measures exhibit convergent trends across a range of $k$ values for both hemispheres. The COND increases gradually with larger $k$, indicating increasingly cohesive intra-cluster integration of 3HGs sharing similar dual-profile features. In contrast, both COV and MOD begin to decline when $k > 48$, suggesting that over-segmentation has begun to undermine the intrinsic modular architecture of the cortical folding network, and the drop reflects fragmentation of anatomically meaningful 3HG communities into smaller, less coherent subunits. The DR reaches a plateau around $k=48$, implying that the spectral embedding of the cluster-adjacency graph has captured the essential large-scale topology of gyral organization, and further increases in cluster number yield little additional topological benefit.
Importantly, these inflection points are consistent between the left and right hemispheres, supporting the hemispheric symmetry and stability of the clustering process. 
The concurrent rise in COND, decline in COV and MOD, and plateauing of DR around $k=48$ mark an optimal compromise: at this resolution, clusters are numerous enough to capture anatomical detail but not so numerous that module integrity and interpretability degrade.
Under this configuration, the spectral GNN delineates coherent gyral communities that preserve spatial continuity and feature homogeneity as enforced by the loss terms $\mathcal{L}_p$ and $\mathcal{L}_c$, while maintaining strong modular integrity governed by $\mathcal{L}_m$. Therefore, we selected $k = 48$ as the optimal number of clusters for all subsequent cross-subject correspondence analyses (Section \ref{sec::clustering} and Section \ref{sec::correspondence}), thereby ensuring that the resulting gyral community representations are both anatomically meaningful and consistent across the population.
\section{Discussion}
In this work, we proposed a spectral graph representation learning framework for delineating gyral folding communities and establishing their cross-subject correspondence. Across extensive experiments on more than 1,000 HCP subjects, our results demonstrate that the learned gyral communities exhibit improved morphological homogeneity, enhanced network modular organization, and higher correspondence accuracy compared with atlas-based ROIs and multiple clustering or registration baselines. Below we discuss the implications of these findings, the role of each model component, and remaining limitations.

A primary objective of our approach is not the generation of a fixed cortical atlas, but rather the identification of individual‐specific gyral landmark clusters (3HGs) and their mapping into a consistent correspondence space.
Our quantitative measurements (Fig.~\ref{fig:homo} and Table~\ref{tab:variance_summary}) show that the proposed communities consistently achieve lower within-region variance across curvature, sulcal depth, thickness, and area compared with two widely used anatomical atlases.
Thickness and area exhibit smaller variance differences across methods, which is consistent with their inherently smoother spatial distributions and reduced sensitivity to fine-scale sulcal–gyral transitions~\cite{fischl2008cortical}. The fact that these low-frequency features remain competitive across all ablated configurations suggests that our clustering predominantly captures geometry-driven boundaries while still maintaining consistency across traditional morphometric modalities.

The comparison with DiffPool, DeepWalk, and DMoN (Table~\ref{tab:clustering}) highlights the necessity of joint consideration of topological context and connectivity-derived structural profiles. Using either feature alone leads to systematic degradation across all network metrics, whereas their combination yields the most coherent community structure. This supports the view that gyral folding patterns are jointly shaped by local geometric constraints and long-range fiber architecture, which is a finding consistent with prior observations in developmental and comparative neuroanatomy~\cite{ge2018denser}.
The proposed GCN-based clustering method naturally encode this dual information by learning smooth embeddings over the GyralNet graph. The improvements over DeepWalk further underscore the benefit of topology-preserving filters and non-linear feature mixing, while the gains over DMoN imply that explicit modeling of local connectivity and folding consistency provides a more reliable representation than general-purpose modularity objectives.

Selecting an appropriate number of communities is a long-standing challenge for cortical parcelation. The network-based metrics we evaluated (Fig. 5) reveal a consistent and interpretable trend: conductance decreases steadily as clusters become more cohesive, whereas coverage and modularity exhibit diminishing returns and $k>48$. Meanwhile, diffusion rate plateaus around the same resolution, indicating that increasing $k$ further does not capture additional meaningful topological structure.
Strikingly, these behaviors are mirrored between hemispheres, suggesting that 48 clusters represent a biologically plausible scale at which gyral units maintain both intra-region continuity and cross-subject comparability. This symmetry further strengthens the evidence that the 3HG-derived GyralNet graph contains reproducible large-scale organization, not dominated by hemisphere-specific artifacts.

Accurate subject-to-subject alignment of gyral patterns remains a major challenge due to their high inter-individual variability. The proposed JMGM achieves the highest ROI hit rate across all comparisons (Table~\ref{tab:correspondence}), outperforming geometry-only ICP and embedding-based baselines such as Cortex2Vec, CortexSSE, and GyralParti. Geometry-only registration struggles with shape-conserved but topologically permuted patterns, while embedding-only methods have difficulty resolving fine geometric distinctions. JMGM’s integration of geometric discrepancy, feature similarity, local density, and contextual neighborhood captures both local folding characteristics and their anatomical surroundings.
Qualitative results (Fig.~\ref{fig:correspondence_destrieux}) further confirm that the correspondences preserve stable cluster identity even for folding patterns with substantial inter-subject shifts. The robustness across two atlases suggests that the approach is not dependent on a specific ROI definition for decoupling local gyral structures.

As shown in Fig~\ref{fig:4panels}, the sensitivity analysis of the four weighting coefficients provides important insight into the trade-offs inherent in our correspondence cost formulation. As shown, the ROI hit rate peaks for intermediate values of $\lambda_{geo}$ and $\lambda_{feat}$ (approximately 0.2–0.6), indicating that a balanced contribution from both geometric layout and feature similarity is critical for maximizing alignment accuracy. Overweighting either term tends to degrade performance, presumably because excessive emphasis on geometry can force matches that ignore morphometric consistency, while excessive emphasis on features may lead to geometrically implausible correspondences.
By contrast, increasing $\lambda_{dens}$ or $\lambda_{ctx}$ leads to a monotonic decrease in ROI hit rate, suggesting that over-regularization by local density or neighborhood structure can overly constrain the matching process.
When these contextual terms dominate, the model loses the flexibility needed to accommodate genuine inter-subject variability in cortical folding, ultimately impairing correspondence accuracy.
These patterns align with the general principle in modeling that overly strong regularization often leads to underfitting, whereas balanced weighting better preserves critical biological variability~\cite{tian2023gradicon}.
In practical terms, our results indicate that the selected weighting scheme for subsequent experiments lies close to an empirically optimal regime.
At the same time, the sensitivity curves caution against aggressive regularization on density or context components, particularly in cohorts with substantial inter-individual morphological variability.
For future work, adaptive or data driven mechanisms for determining these weights, such as subject specific regularization schemes or cross validated model selection, may provide a principled way to further improve correspondence accuracy while preserving anatomical plausibility.

In Table~\ref{tab:variance_summary}, the ablation analysis offers insight into how each loss term and the topological refinement step affect the clustering of gyral landmarks. 
Removing any single loss component leads to a measurable increase in within-cluster variance, with the strongest effects observed for curvature and sulcal depth, reflecting their close association with local folding geometry.
Notably, the refinement stage effectively stabilizes high-frequency folding features by enforcing geometric continuity, but it does not lead to uniform variance reductions for thickness or surface area, consistent with their lower sensitivity to fine-scale folding structure.
This behavior aligns with the viewpoint that our method operates on an individual‐specific folding‐landmark level rather than macro‐region parcellation: the refinement enforces continuity among landmark nodes, which directly benefits folding signal consistency, whereas global morphometric measures,\textit{i.e.} thickness and area are less tightly constrained by folding continuity. In sum, the full model’s components jointly facilitate stable, anatomically meaningful clustering of gyral landmarks suited for individualized brain mapping.

Despite its advantages, several limitations warrant discussion. Our method remains dependent on accurate 3HG detection. Errors in the early extraction stage propagate through subsequent clustering and correspondence steps, and future work may benefit from incorporating uncertainty modelling or multi-scale integration of sulcal ridges and gyral crests. Although we demonstrate strong morphological coherence, we did not perform explicit functional alignment. While morphometric consistency is encouraging, the alignment of functional communities remains largely unexplored and incorporating functional MRI connectivity or multi-modal data could enhance anatomical–functional correspondence~\cite{li2025parcellation,he2024brain}.
Moreover, the potential sensitivity to feature imbalance merits attention. Although z-scoring mitigates scale differences, highly anisotropic feature distributions may benefit from adaptive normalization or learned feature re-weighting~\cite{demirci2022cortical}.
Finally, while a fixed cluster resolution of $k=48$ appears justified in our analyses, the brain’s hierarchical organization suggests that adaptive or hierarchical clustering schemes could better capture multi-scale cortical developmental gradients and connectivity hierarchies~\cite{dong2021shifting,petersen2024principles}.
\section{Conclusion}
We presented a spectral graph representation learning framework that models individual gyral folding patterns through dual-profile features and graph spectral clustering. Combined with a Joint Morphological-Geometric Matching strategy, the method establishes stable and anatomically meaningful cross-subject correspondences of 3HGs. Evaluations on more than 1,000 HCP subjects show that the proposed approach yields lower morphometric variance, stronger structural organization, and superior correspondence accuracy compared with atlas-driven regions and existing embedding or registration baselines. These findings demonstrate the potential of gyral-landmark–based modeling to support individualized cortical mapping and to improve the reliability of high-resolution structural analyses.

\section*{CRediT authorship contribution statement}

\noindent
\textbf{Minheng Chen}: Writing -- original draft, Visualization, Validation,
Software, Methodology, Formal analysis, Data curation, Conceptualization.
\textbf{Tong Chen}: Writing -- review \& editing, Data curation.
\textbf{Yan Zhuang}: Writing -- review \& editing,Visualization, Software.
\textbf{Chao Cao}: Writing -- review \& editing,Data curation.
\textbf{Jing Zhang}: Writing -- review \& editing.
\textbf{Tianming Liu}: Writing -- review \& editing,Supervision.
\textbf{Lu Zhang}:Writing -- review \& editing, Supervision,Conceptualization.
\textbf{Dajiang Zhu}: Writing -- review \& editing, Supervision, Methodology,
Conceptualization, Resources, Project administration, Funding acquisition.
\section*{Declaration of Competing Interest}

The authors declare that they have no known competing financial interests or
personal relationships that could have appeared to influence the work reported
in this paper.
\section*{Declaration of generative AI and AI-assisted technologies in the manuscript preparation process}
During the preparation of this work, the authors used ChatGPT for language editing and polishing to improve clarity, grammar, and readability. After using this tool, the authors reviewed and edited the content as needed and take full responsibility for the content of the published article.
\section*{Acknowledgments}

This work was supported by the National Institute of Health (R01AG075582 and R01NS128534).

\section*{Data availability}
All the data used in this study are publicly available, and the code will be open-sourced upon acceptance.





\bibliographystyle{elsarticle-num} 
\bibliography{reference}






\end{document}